%% file: main.tex
\documentclass[letterpaper,journal]{IEEEtran}
\pdfoutput=1
\usepackage{array}
\usepackage{cite}
\usepackage{textcomp}
\usepackage{stfloats}
\usepackage{url}
\usepackage{verbatim}
\usepackage{balance}

\usepackage[cmex10]{amsmath}
\usepackage{amsthm}
\usepackage{amssymb}
\usepackage{braket}
\usepackage[short]{optidef}

\usepackage{graphicx}
\usepackage[caption=false, font=normalsize, labelfont=sf, textfont=sf]{subfig}

\theoremstyle{plain}
\newtheorem{theorem}{Theorem}
\newtheorem{proposition}{Proposition}
\newtheorem{lemma}{Lemma}

\newcommand{\bbR}{\mathbb{R}}
\renewcommand{\epsilon}{\varepsilon}
\renewcommand{\theta}{\vartheta}
\renewcommand{\rho}{\varrho}
\renewcommand{\phi}{\varphi}

\newcommand{\prob}[1]{p_{#1}}

\newcommand{\norm}[1]{\lVert #1 \rVert}

\newcommand{\deriv}[2]{\partial_{#2 } #1}

\newcommand{\PhiBA}{\Phi_{\text{BA}}}

\newcommand{\supp}[1]{\text{supp}( #1 )} 
\newcommand{\supporto}{E}
\newcommand{\relint}[1]{\mathrm{int}{( #1 )}}
\renewcommand{\vec}[1]{\boldsymbol{#1}}
\newcommand{\mat}[1]{\boldsymbol{#1}}

\newcommand{\myparagraph}[1]{\noindent\textbf{#1.}}

\usepackage{hyperref}

\begin{document}
\title{Vector Flows and the Capacity of a Discrete Memoryless Channel}

\author{
Guglielmo Beretta, Giacomo Chiarot, Antonio Emanuele Cin\`{a}, and Marcello Pelillo, \IEEEmembership{Fellow, IEEE}%
\thanks{This work has been submitted to the IEEE for possible publication. Copyright may be transferred without notice, after which this version may no longer be accessible.}
\thanks{G.\ Beretta is with Ca' Foscari University of Venice, 30170 Venice, Italy, and also with Polytechnic University of Turin, 10138 Turin, Italy.}
\thanks{A.\ E.\ Cin\`{a} is with University of Genoa, 16145 Genoa, Italy.}
\thanks{G. Chiarot and M.\ Pelillo M.~Pelillo are with Ca' Foscari University of Venice, 30170 Venice, Italy.}
}
\maketitle
\begin{abstract}
\input{sections/abstract}
\end{abstract}
\begin{IEEEkeywords}
    Analog computation, capacity, convex optimization, discrete memoryless channel,  dynamical systems, mutual information, ODE, optimal input distribution,  vector flow.
\end{IEEEkeywords}
\IEEEpeerreviewmaketitle
\section{Introduction}\label{sec:introduction}
\input{sections/introduction}
\input{sections/notation}
\input{sections/channel_classical}
\input{sections/general_flows}
\input{sections/flow_for_i}
\input{sections/experiments}
\input{sections/results}
\input{sections/discussion}
\input{sections/conclusion}

\appendices

\section{Proofs and computations}\label{app:proofs}
\input{sections/proofs}
\section{Remarks about the modified Euler method}\label{app:odeNorm}
\input{sections/implementation}

\subsection*{Acknowledgment}
    Guglielmo Beretta's scholarship is jointly funded by Polytechnic University of Turin and Ca' Foscari University of Venice.
    The authors thank Sebastiano Vascon for the helpful conversations about this manuscript. 
\bibliographystyle{ieeetr}
\bibliography{refs}
\end{document}

%% file: sections/abstract.tex
One of the fundamental problems of information theory, since its foundation by Shannon in 1948, has been the computation of the capacity of a discrete memoryless channel, a quantity expressing the maximum rate at which information can travel through the channel.
In the literature, several algorithms were proposed to approximately compute the capacity of a discrete memoryless channel, being an analytical solution unavailable for the general discrete memoryless channel. This paper presents a novel approach to compute the capacity, which is based on a continuous-time dynamical system.
Such a dynamical system can indeed be regarded as a continuous-time version of the Blahut-Arimoto algorithm. In fact, the updating map appearing in the Blahut-Arimoto algorithm is here obtained as a suitable discretization of the vector flow presented, using an analogy with some game-theoretical models.
Finally, this analogy suggests a high-level hardware circuit design enabling analog computation to estimate the capacity.

%% file: sections/introduction.tex
\IEEEPARstart{E}{stimating} the capacity of discrete memoryless channel (DMC) is of main importance in quantifying the reliability of point-to-point communication systems as a consequence of Shannon's noisy-channel coding theorem \cite{shannon_1948_a_mathematical,mackay_information_2003,venkatesan_2020_10.1145/3357713.3384323}.
Among the algorithms devised to compute the capacity of a DMC, a fundamental result is the classical Blahut-Arimoto algorithm (BAA), an iterative algorithm based on an alternating maximization procedure \cite{csiszar_1984_information_geometry}. The BAA, published in 1972 and named after S.~Arimoto \cite{arimoto:capacity} and R.~Blahut \cite{blahut:capacity}, who discovered it independently, allowed for the first time to deal with DMCs operating on input and output alphabets of different size, requiring just some mild conditions on the zero elements of the transition matrix.

Several algorithms in the literature have been derived from the BAA. Those works adopt diverse techniques to improve the BAA, \textit{e.g.}, speeding it up using information geometry concepts \cite{Matz_2004_1405276} or adopting squeezing strategies \cite{yu_2010_5484972}. The BAA is still a subject of active research, see \textit{e.g.}  the recent article by Boche \textit{et al.} \cite{boche_2023_10130754}, disclosing a remarkable limitation of the Turing machine model when it comes to compute an optimal input distribution for a DMC.

It is worth mentioning that there exist alternative approaches to estimate the capacity that are substantially different from the BAA, see \textit{e.g.}  \cite{sutter_2017_thesis_20.500.11850/218720}, based on convex optimization, or the recent work by Tope and Morris \cite{tope_2021_9400323}, where statistical inference is performed to overcome the lack of information about the transition matrix.

This paper presents a novel approach to compute the capacity of a DMC. First, we recast the problem into a constrained optimization program on a standard simplex. We then introduce a game-theoretical inspired continuous-time dynamical system~\cite{hofbauer_evolutionary_1998} defined through an ordinary differential equation (ODE), and we finally show that the proposed dynamic can be used to solve the aforementioned optimization program, as it yields a first-order interior-point method~\cite{luenberger_linear_2016}.

Notably, ODEs and the associated vector flows have been widely used for constrained optimization, see \textit{e.g.}  \cite{helmke_optimization_1994} and the references therein. However, to the best of our knowledge, no previous work applies vector flows to compute the capacity of DMCs.

Interestingly, this kind of ODEs also appeared in
\cite{brockett_dynamical_1988} as a way to tackle some linear algebra problems by casting them into suitable optimization programs. The flow presented in \cite{brockett_dynamical_1988} is an instance of the Toda flow, a matrix flow related to optimization problems on certain submanifolds of matrix spaces \cite{Chu_1988_isospectral_flows}. A somewhat more general formulation than the flow of \cite{brockett_dynamical_1988} has been then discussed
in \cite{Faybusovich1991DynamicalSW}, where smooth functions are optimized over polyhedra using some notions of Riemannian geometry. Indeed, we realized that our method can be regarded as an instance thereof in case the objective function is smooth up to the boundary of the feasible set. However, in contrast to \cite{Faybusovich1991DynamicalSW}, we shall see that our objective function may lack differentiability at the boundary of the feasible set.

Thanks to the analogy of our method with the replicator dynamics, which is a model pertaining evolutionary game theory \cite{hofbauer_evolutionary_1998}, it is here shown that the updating map appearing in the BAA can be recovered by discretizing the flow associated with the ODE that we describe. This motivates the claim that the method proposed can be regarded as a continuous-time version of the BAA.
Moreover, in case the objective function is a (possibly higher-order) polynomial, the flow here discussed can be regarded as the continuous-time counterpart of the Baum-Eagon map \cite{baum_inequality_1967,baum_growth_1968} which has recently been generalized in \cite{palaiopanos_multiplicative_2017,panageas_multiplicative_2019}.

The subsequent sections of this paper are organized as follows. We commence by establishing notation conventions in Section~\ref{sec:notation}.
In Section~\ref{sec:problem}, we rigorously formalize the challenge of computing channel capacity for DMCs.
The crux of our contribution lies in Section~\ref{sec:general-flows}, where we introduce the vector flow optimization framework. This framework is further refined in Section~\ref{subsec:flow-for-i}, where the vector flow allowing to compute the capacity of a DMC is formulated.
Our experimental design and setup are explicated in Section~\ref{sec:experiments}, where we elucidate relevant preliminaries, articulate primary objectives, and present the empirical findings. 
Subsequently, Section~\ref{sec:discussion} summarizes the key outcomes derived from the theoretical and experimental analysis.
Finally, we expound upon the contributions and delineate potential avenues for future research in Section~\ref{sec:conclusion}. Noteworthy proofs of our theorems and lemmas are predominantly situated in Section~\ref{sec:general-flows} and Section~\ref{subsec:flow-for-i}, with supplementary proofs relegated to Appendix~\ref{app:proofs} and Appendix~\ref{app:odeNorm}.

%% file: sections/notation.tex
\section{Notations}\label{sec:notation}
\noindent In this paper, the information content is tacitly measured in \emph{nats}.
This choice simplifies the computations, as mentioned, \textit{e.g.}, in \cite{blahut:capacity}. For a conversion to bits, we recall that $1$ nat equals $\ln{2}$  bits \cite{mackay_information_2003}.

For a discrete random variable $X$ with range in the set $\mathcal{X}$ we set $\prob{X}(x) = \text{prob}(X = x)$ for every $x \in X$. 
For every positive integer $n$, the \emph{standard simplex} in $\bbR^{n}$ is the set
\begin{equation*}
    \Delta_{n} = \Set{ \Vec{z} \in \bbR^{n} | \Vec{z}\geq \Vec{0}\quad\text{and}\quad\sum_{i = 1}^{n} z_i = 1},
\end{equation*}
its (relative) \emph{interior} is the set
\begin{equation*}
    \relint{\Delta_{n}} = \Set{ \Vec{z} \in \bbR^{n} | \Vec{z} > \Vec{0}\quad \text{and}\quad \sum_{i = 1}^{n} z_i = 1},
\end{equation*}
and its (relative) \emph{boundary} is the set $\partial{\Delta_{n}} = \Delta_{n} \setminus \relint{\Delta_{n}}$.
Furthermore, for $\Vec{z}=(z_1, \dots, z_n) \in \bbR^n$ we call \emph{support} of $\Vec{z}$ the set $\supp{\Vec{z}}=\set{i \in [n] | z_i \neq 0}$.
We use $\deriv{}{i}$ as an alias for the derivation $\partial/ \partial z_i$, and for a function $\Vec{z}(t)$ depending on the variable $t$, the function $\dot{\Vec{z}} = \dot{\Vec{z}}(t)$ denotes the derivative of $\Vec{z}(t)$ with respect to $t$.

%% file: sections/channel_classical.tex
\section{Problem formulation}\label{sec:problem}
\noindent A discrete memoryless channel (DMC) is a communication system described by a triplet $(\mathcal{X}, \mathcal{Y}, \mat{P})$, in which $\mathcal{X}= \{x_1, x_2, \dots, x_n\}$ and  $\mathcal{Y} = \{ y_1, y_2, \dots,  y_m\}$ are finite alphabets called \emph{input alphabet} and \emph{output alphabet} respectively, whereas $\mat{P} =[p(j|i)]_{i \in [n], j \in [m]}\in \bbR^{n \times m}$ is a stochastic matrix, called \emph{transition matrix}, where $p(j|i)$ expresses the probability that the symbol $y_j$ is observed as output of the system whenever the symbol $x_i$ is sent to the system as input \cite{cover_elements_2006}.
The transition matrix determines the relation existing between the \emph{input variable} and the \emph{output variable} of the channel, \textit{i.e.}, between the random variable $X$ with range in $\mathcal{X}$ that models the input of the channel, and the random variable $Y$ with range in $\mathcal{Y}$ that models the corresponding output of the channel.
We shall assume that, for every $j \in [m]$, there exists at least one $i \in [n]$ such that $p(j|i) > 0$. This property aligns with the notion that $\mathcal{Y}$ represents the minimal output alphabet required for a description of the DMC. In fact, this means that for any selected symbol $y \in \mathcal{Y}$, there exists a corresponding input distribution for which $y$ occurs as output with positive probability.%
    \footnote{This assumption is not restrictive. In the general scenario, it is sufficient to consider a smaller output alphabet, obtained by removing from $\mathcal{Y}$ the symbols $y_j$ such that $p(j|i) = 0$ for every $i \in [n]$.}
The information content that can be conveyed through the channel is captured by the \emph{mutual information} between the input variable $X$ and the output variable $Y$, denoted as $I[X;Y]$. As mentioned in \cite{mackay_information_2003}, there are several equivalent formulations for the mutual information of $X$ and $Y$. Among them, we shall consider the following:
\begin{equation*}
    I[X;Y] = H[Y] - H[Y|X]
\end{equation*}
where $H[Y]$ is the \emph{entropy}%
    \footnote{Here and in the sequel, we adopt the usual convention that $\alpha \ln{\alpha} = 0$ for $\alpha = 0$.}
 of $Y$:
\begin{equation*}
    H[Y] = -\sum_{y \in\mathcal{Y}} \prob{Y}(y)\ln[\prob{Y}(y)]
\end{equation*}
and $H[Y|X]$ is the \emph{conditional entropy} of $Y$ given $X$:
\begin{equation*}
    H[Y|X] = 
    \sum_{x\in \mathcal{X}}
    \prob{X}(x)
    H[Y|X=x],
\end{equation*}
being
\begin{equation*}
    H[Y|X=x] = 
    -\sum_{y \in \mathcal{Y}}
    \prob{Y|X=x}(y)
    \ln[\prob{Y|X=x}(y)]
\end{equation*}
for every $x \in \mathcal{X}$.
The \emph{capacity} of the channel is defined as
\begin{equation*}
    C = \max_{X} I[X; Y],
\end{equation*}
\textit{i.e.}, the maximum of the mutual information among all possible input and corresponding output variables.
The channel capacity is central in Shannon's noisy-channel coding theorem, a cornerstone of information theory. This theorem establishes a fundamental connection between the channel capacity and the rate of lossless communication across the channel. 
Moreover, using appropriately an encoder and a decoder, a message can be conveyed through the channel at a rate that is arbitrarily close to the theoretical bound, as elucidated in \cite{shannon_1948_a_mathematical, mackay_information_2003}. In particular, polar codes \cite{Arikan_2008_channel_polarization} provide a principled way to approach this bound in case the input alphabet is binary.

%% file: sections/general_flows.tex
\section{Maximizing vector flows}\label{sec:general-flows}
\subsection{Regular flows}\label{subsec:regular-flows}
\noindent We now delve into a more comprehensive construction to solve a program of the form
\begin{maxi}|s|[0]
    {\Vec{z} \in \Delta_{n}} 
    {f(\Vec{z})} 
    {\label{p:f}}
    {} 
\end{maxi}
in which the objective function $f$ is continuous on $\Delta_n$ and differentiable in $\relint{\Delta_n}$.
This serves our purpose, since the problem of estimating the capacity of a DMC can be recast into an optimization program of this type.
The outcome of the construction we are about to discuss is a continuous-time first-order method for constrained optimization on $\Delta_n$.
This method relies on a vector flow that, as we will demonstrate, arises from an ODE%
    \footnote{A more general class of ODEs can be derived using ideas from Riemannian geometry, see \textit{e.g.} \cite{Faybusovich1991DynamicalSW}.}
depending on the gradient of $f$.

To simplify the exposition, we commence by considering a simplified case, in which the objective function, as we shall see, is smooth up to the boundary of the feasible set, with derivatives that are Lipschitz-continuous.
Let $\Omega$ be an open subset of $\bbR^{n}$, suppose that $f \in C^{1,1}(\Omega)$, \textit{i.e.}, that $f$ is a real-valued function defined on $\Omega$ that admits a Lipschitz-continuous gradient in $\Omega$, and define the ODE%
    \footnote{The reader may notice that \eqref{eq:ode-grad}, in case the function $f$ is a quadratic form, coincides with  the classical replicator dynamics \cite{weibull_evolutionary_1995}.}
describing the evolution of $\Vec{z} \in \Omega$:
\begin{align}\label{eq:ode-grad}
    \dot{z}_{i}
    &=
    z_{i}
    \left[ \deriv{f}{i}(\Vec{z}) 
    - \sum_{k = 1}^n z_{k } \deriv{f}{k}(\Vec{z}) \right], & i \in [n].
\end{align}

It is convenient to introduce $\vec{g}=(g_1, \dots, g_n) \colon \Omega \to \bbR^{n}$, defined for every $i \in [n]$ and every $\vec{z}\in \Omega$ by
\begin{equation*}
    g_i(\vec{z}) = \deriv{f}{i}(\Vec{z}) 
    - \sum_{k = 1}^n z_{k } \deriv{f}{k}(\Vec{z})
\end{equation*}
in such a way that that \eqref{eq:ode-grad} can be rewritten as
\begin{align*}    
	\dot{z}_{i} &=  {z}_{i} g_i(\vec{z}), & i \in [n].
\end{align*}
Equivalently, \eqref{eq:ode-grad} can be written as
\begin{equation}\label{eq:ode-g}
	\dot{\vec{z}} = \vec{z} \odot \vec{g}(\vec{z})
\end{equation}
where $\odot$ denotes the Hadamard (element-wise) product between vectors.

Under the assumption $\Delta_n \subset \Omega$, which we make throughout all Section~\ref{sec:general-flows}, it is straightforward to check that \eqref{eq:ode-g} is \emph{regular} in the sense of \cite{weibull_evolutionary_1995}, \textit{i.e.}, that
\begin{equation*}
    \sum_{i = 1}^n z_i g_i(\vec{z}) = 0
\end{equation*}
for every $\vec{z} \in \Delta_{n}$.
Thanks to the regularity condition, the set $\Delta_{n}$ results invariant under the dynamics \cite{weibull_evolutionary_1995}.
\begin{theorem}[Dynamics on the standard simplex \cite{weibull_evolutionary_1995}]\label{thm:invariant}
    For every $\tilde{\Vec{z}} \in \Delta_n$ there exists one and only one differentiable function $\Vec{z} = \vec{z}(t)$ defined for $t \in [0, +\infty)$ that satisfies
    \begin{equation}\label{eq:ivp}
        \begin{cases}        
            \dot{z}_{i} &= z_{i} \left[ \deriv{f}{i}(\Vec{z}) 
        - \sum_{k = 1}^n z_{k } \deriv{f}{k}(\Vec{z}) \right],  \quad i \in [n]
        \\
        \Vec{z}(0) &= \tilde{\Vec{z}}.
        \end{cases}
    \end{equation}
    Furthermore, $\vec{z}(t) \in \Delta_{n}$ and $\supp{\vec{z}(t)} = \supp{\tilde{\Vec{z}}}$ for every $t \in [0, +\infty)$.
\end{theorem}
See \cite{weibull_evolutionary_1995} for a proof based on the Picard-Lindel\"{o}f's theorem \cite{birkhoff_rota_ODE}, which is applicable due to the hypothesis $f \in C^{1, 1}(\Delta_n)$.
    The underlying idea is to show that the hyperplanes
    \begin{equation*}
        H_0 = \set{\vec{z} \in \bbR^{n} | \sum_{i = 1}^n z_i = 1}
    \end{equation*}
    and
    \begin{align*}
        H_i &= \set{\vec{z} \in \bbR^{n} | z_i = 0}, & i\in [n],
    \end{align*}
    are invariant under the flow defined on $\Delta_n$.
    
For $\tilde{\Vec{z}} \in \Delta_n$, we refer to the solution $\Vec{z} = \Vec{z}(t)$ to \eqref{eq:ivp} defined on the interval $[0, +\infty)$ as the (forward) \emph{trajectory of \eqref{eq:ode-grad} initialized in $\tilde{\Vec{z}}$}. The main feature of \eqref{eq:ode-grad} is that the function $f$ increases strictly along every non-constant trajectory of the vector flow defined by \eqref{eq:ode-grad}. According to the terminology pertaining evolutionary game theory,
this can be expressed by saying that $f$ is a \emph{strict Lyapunov function} \cite{helmke_optimization_1994} for \eqref{eq:ode-grad}.
\begin{theorem}[Lyapunov function]\label{thm:lyapunov}
    Every trajectory $\vec{z}(t)$ of \eqref{eq:ode-grad} satisfies $d (f \circ \vec{z}) / dt \geq 0$. Equality holds if and only if $\vec{z}(t)$ is a constant trajectory.
\end{theorem}
\begin{IEEEproof}
    The proof proceeds along the same line of \cite[Thm.~1]{Bomze1997EvolutionTT}. Let $\vec{z} = \vec{z}(t)$ be a trajectory of \eqref{eq:ode-grad} and define $\mu = \mu(t)$ by $\mu = \sum_{k = 1}^n z_{k} \deriv{f}{k}(\Vec{z})$.
    Since $\sum_{i = 1}^{n} \dot{z}_{i} = 0$ by the regularity condition, this means that
    \begin{align*}
        \dfrac{d}{dt} (f \circ \vec{z})
        &= \sum_{i = 1}^{n} \dot{z}_{i}  \deriv{f}{i}(\vec{z})\\
        &= \sum_{i = 1}^{n} \dot{z}_{i} ( \deriv{f}{i}(\vec{z}) - \mu)\\
        &= \sum_{i = 1}^{n} {z}_{i} ( \deriv{f}{i}(\vec{z}) - \mu)^2,
    \end{align*}
    which is a sum of non-negative quantities. Consequently, the derivative computed is non-negative, and it is equal to zero for some $t \in \bbR$ if and only if ${z}_{i}(t) [ \deriv{f}{i}(\vec{z}(t)) - \mu(t)]= 0$ for every $i \in [n]$, \textit{i.e.}, if and only if $\dot{\vec{z}}(t) = \vec{0}$. Observe now that $\dot{\vec{z}}(t) = \vec{0}$ for some $t$ if and only if the trajectory is constant, once again due to the Picard-Lindel\"{o}f's theorem.
\end{IEEEproof}

The content of Theorem~\ref{thm:lyapunov} is also discussed in \cite{Faybusovich1991DynamicalSW}, where it motivates the use of the ODE considered as a continuous interior-point method \cite{luenberger_linear_2016}, and this is precisely what we are interested in.
The fact that the dynamics admits a Lyapunov function has a well-known consequence, which rules out chaotic behaviour and limit cycles \cite{hofbauer_evolutionary_1998,weibull_evolutionary_1995}. 

We recall first that $\vec{z} \in \Delta_{n}$ is a \emph{stationary point} for \eqref{eq:ode-grad} if $\vec{z} \odot \vec{g}(\vec{z})= \vec{0}$.  
\begin{theorem}[Trajectories converge \cite{hofbauer_evolutionary_1998}]\label{thm:non-chaos}
    Every trajectory $\vec{z}(t)$ of \eqref{eq:ode-grad} initialized in $\Delta_n$ admits a limit in $\Delta_{n}$ for $t \to +\infty$, which is a stationary point for \eqref{eq:ode-grad}.
\end{theorem}
It is straightforward to see that $\vec{z} \odot \vec{g}(\vec{z})= \vec{0}$ if and only if
\begin{equation}\label{eq:fix}
    \deriv{f}{i}(\vec{z})
    = \sum_{k = 1}^n z_{k} \deriv{f}{k}(\vec{z}) \quad \text{for every $i \in \supp{\vec{z}}$}
\end{equation}
and so $\Vec{z}\in \Delta_n$ is a stationary point for \eqref{eq:ode-grad} if and only if it satisfies \eqref{eq:fix}.
Moreover, \eqref{eq:fix} is valid in case a stronger condition holds, the Karush-Kuhn-Tucker (KKT) condition --- also known as first-order optimality condition \cite{luenberger_linear_2016} --- for \eqref{p:f}. This means in particular that if $\Vec{z} \in \Delta_n$ is a KKT point for \eqref{p:f}, then $\Vec{z}$ is a stationary point for \eqref{eq:ode-grad} --- see \textit{e.g.} Proposition~\ref{prop:kkt-equivalence} in Appendix~\ref{app:proofs}.
Furthermore, there exists a partial converse to this implication. Specifically, a stationary point for \eqref{eq:ode-grad} is also a KKT point for \eqref{p:f} if it is the limit point of a trajectory of \eqref{eq:ode-grad} that is suitably initialized. 
\begin{proposition}[Limit points]\label{prop:limit}
Let $\vec{z}(t)$ be a trajectory of \eqref{eq:ode-grad} and set $\vec{z}^{*} = \lim_{t \to +\infty}\vec{z}(t)$. If $\vec{z}(0) \in \relint{\Delta_{ k}}$, then $\vec{z}^{*}$ is a KKT point for \eqref{p:f}.
\end{proposition}
\begin{IEEEproof}
    Set $C= \sum_{i = 1}^n z_{ i}^{*} \deriv{f}{ i}(\vec{z}^{*})$.  
    The point $\vec{z}^{*}$ is a limit point of a trajectory of \eqref{eq:ode-grad} and so it is a stationary point.
    By \eqref{eq:fix}, the equality $\deriv{f}{i}(\vec{z}^{*})
    = C$ holds for every $i \in \supp{\vec{z}^{*}}$, whereas, by Lemma~\ref{lem:mult}, the inequality $\deriv{f}{i}(\Vec{z})^{*} \leq C$ holds for every $i \in [n] \setminus \supp{\vec{z}^{*}}$.
    By Proposition~\ref{prop:kkt-equivalence}, it follows that $\vec{z}^{*}$ is a KKT point for \eqref{p:f}.
\end{IEEEproof}

The results discussed so far culminate in Theorem~\ref{thm:concave}, which establishes a method to solve \eqref{p:f} when it is a concave program. Under this additional concavity assumption, a global solution to \eqref{p:f} may be found as the limit point of any trajectory of \eqref{eq:ode-grad} initialized within the relative interior of $\Delta_{n}$. Following this trajectory is henceforth a principled approach for tackling \eqref{p:f}.
\begin{theorem}[Concave maximization]\label{thm:concave}
Let $\vec{z}(t)$ be a trajectory of \eqref{eq:ode-grad} initialized in $\Vec{z}(0) \in \relint{\Delta_n}$ and suppose $f$ is a concave function.
Then $\lim_{t \to +\infty}\vec{z}(t) \in \arg \max_{\vec{z} \in \Delta_n} f(\vec{z})$.
\end{theorem}
\begin{IEEEproof}
    By Theorem~\ref{thm:non-chaos},  the trajectory $\vec{z}(t)$ admits a limit as $t \to +\infty$, which is a KKT point for \eqref{p:f} as a consequence to Proposition~\ref{prop:limit}. By hypothesis, $f$ is concave and differentiable up to the boundary, hence being a KKT point for \eqref{p:f} is equivalent to being a global solution to \eqref{p:f}.
\end{IEEEproof}

We remark that, among the hypothesis of Theorem~\ref{thm:concave}, the condition on the trajectory initialization is fundamental. Specifically, for a trajectory that is initialized on $\partial{\Delta_{n}}$, the flow is generally unable to guide towards a global solution for \eqref{p:f}.
In fact, Theorem~\ref{thm:invariant} implies that the support of the starting point of the dynamics is a superset of the support of the limit point.
However, if $f$ is concave, then the dynamics is still solving a maximization program for which $f$ is the objective function. This time, however, the optimization takes place on a feasible set that is different from $\Delta_n$.
The new feasible set is a subset of $\Delta_n$ that incorporates an additional support constraint,
as elucidated in Theorem~\ref{thm:concave-face}.
\begin{theorem}[Concave maximization with support constraint]\label{thm:concave-face}
Let $\vec{z}(t)$ be a trajectory of \eqref{eq:ode-grad} and suppose $f$ is a concave function.
Then $\lim_{t \to +\infty}\vec{z}(t) \in \arg \max_{\vec{z} \in S} f(\vec{z})$, where
$S = \set{ \vec{z} \in \Delta_{ k} \mid \supp{\vec{z}} \subseteq \supp{\vec{z}(0)}}$.
\end{theorem}
\begin{IEEEproof}
    The trajectory $\vec{z}(t)$ admits a limit as $t \to +\infty$ by Theorem~\ref{thm:non-chaos}. 
    Denote such limit by $\vec{z}^{*}$ and set $\supporto_{\infty}= \supp{\vec{z}^{*}}$ and $\supporto_t= \supp{\vec{z}(t)}$ for every $t \geq 0$. Thanks to the continuity of the flow and the uniqueness of solutions, $\supporto_{\infty} \subseteq \supporto_{t} = \supporto_{0}$ for every $t \geq 0 $. It would then be sufficient to observe that that Slater condition holds for the maximization program restricted to $S$, together with Lemma~\ref{lem:mult}, which entails that $\vec{z}^{*}$ satisfies the KKT conditions for the restricted program \cite{boyd_vandenberghe_2004}. We offer here an alternative direct proof. 
    What we have deduced so far proves that $f(\vec{z}^{*}) \leq \max_{\vec{z} \in S} f(\vec{z})$. For the reverse inequality, observe first that $\vec{z}^{*}$ is a stationary point, and so there exists a real constant $C$ such that $\deriv{f}{i}(\vec{z}) = C$ for every $i \in \supporto_{\infty}$. By Lemma~\ref{lem:mult}, the inequality $\deriv{f}{i}(\vec{z}) \leq C$ holds for every $i \in \supporto_{0} \setminus \supporto_{\infty}$.
    The function $f$ is concave, thus the inequality $f(\vec{z}) \leq f(\vec{z}^{*}) + \sum_{i = 1}^n \deriv{f}{i}(\vec{z}^{*})(z_i - z_i^{*})$ holds for every $\vec{z} \in \Delta_{n}$ (see \textit{e.g.} \cite{boyd_vandenberghe_2004}), which specialized for $\vec{z} \in \supporto_0$ yields:
    \begin{align*}
        &f(\vec{z}) - f(\vec{z}^{*}) \leq \sum_{i = 1}^n \deriv{f}{i}(\vec{z}^{*})(z_i - z_i^{*})\\
        &\qquad= \sum_{ i \in \supporto_0} \deriv{f}{ i}(\vec{z}^{*}) (z_{ i} - z_{ i}^{*})\\
        &\qquad= \sum_{ i \in \supporto_{\infty}} C (z_{ i} - z_{ i}^{*}) +
        \sum_{i \in \supporto_0 \setminus \supporto_{\infty}} \deriv{f}{ i}(\vec{z}^{*}) (z_{ i} - z_{ i}^{*})\\
        &\qquad\leq \sum_{ i \in \supporto_{\infty}} C (z_{ i} - z_{ i}^{*}) +
        \sum_{ i \in \supporto_0 \setminus \supporto_{\infty}} C (z_{ i} - z_{ i}^{*})\\
        &\qquad= C \sum_{i = 1}^n (z_{ i} - z_{ i}^{*})\\
        &\qquad= C(1 - 1) = 0,
    \end{align*}
    hence proving that $f(\vec{z})\leq f(\vec{z}^{*})$.
\end{IEEEproof}

\subsection{Flow discretization}
\noindent The vector flow introduced offers a pathway for addressing \eqref{p:f} by tracing trajectories initialized in $\relint{\Delta_n}$. As shown in Section~\ref{subsec:regular-flows}, if $f$ is concave, then these trajectories converge to global solutions of \eqref{p:f}. However, such trajectories lack of a general analytic solution. In such a scenario, the conventional approach to simulate the dynamics involves employing ODE solvers to discretize the trajectories, as we elaborate in Section~\ref{sec:experiments}.
Besides this approach, we present here an alternative discretization method, which we derived by adapting a technique from evolutionary game theory to our dynamics --- see \textit{e.g.} \cite{weibull_evolutionary_1995}, in which it is used in the deduction of the discrete replicator-dynamics model. As we shall see, the resulting discrete dynamical system satisfies a multiplicative-weight-update rule \cite{Arora2012TheMW}.
\begin{proposition}\label{prop:change_of_variables}
    Let $\Vec{z}^{(0)} \in \Delta_n$ and assume that the function $\vec{z}(t)$ on the interval $[0, T]$ is the unique solution to the initial value problem
    \begin{equation}\label{eq:ratio-ivp}
    \begin{cases}
        \dot{z}_i &= z_{i} \left[ \deriv{f}{i}(\Vec{z}) - \sum_{j = 1}^n z_j \deriv{f}{j}(\Vec{z})\right], \quad i \in [n]\\
        \Vec{z}(0) &= \Vec{z}^{(0)}
    \end{cases}
    \end{equation}
    Let $\Vec{p}(t) = (p_1(t), \dots, p_n(t))$ solve on the same interval $[0, T]$ the initial value problem
    \begin{equation}\label{eq:pop-ivp}
    \begin{cases}
        \dot{p}_i &= p_i \deriv{f}{i}\left(\left(\sum_{i = 1}^n p_i\right)^{-1} \Vec{p}\right), \quad i \in [n]\\
        \Vec{p}(0) &= P_0 \Vec{z}^{(0)}        
    \end{cases}
    \end{equation}
    for some constant $P_0 > 0$. Then $\Vec{z}(t) = \left(\sum_{i = 1}^n p_i(t)\right)^{-1} \Vec{p}(t)$ for every $t \in [0, T]$.
\end{proposition}
\begin{IEEEproof}
    Define the function $\Vec{y}(t) = (\sum_{i = 1}^n p_i(t))^{-1} \Vec{p}(t)$ for every $t \in [0, T]$. It is sufficient to check that $\Vec{y}(t)$ solves \eqref{eq:ratio-ivp} on the interval $[0, T]$. By the uniqueness hypothesis on $\Vec{z}(t)$, this entails $\vec{z}(t) = \Vec{y}(t)$ for every $t \in [0, T]$.
\end{IEEEproof}
Note that in Theorem~\ref{prop:change_of_variables} the quantity $\sum_{i = 1}^n p_i(t)$ is positive by hypothesis for every $t \in [0, T]$, whereas a consequence of the Theorem~\ref{prop:change_of_variables}  is that $\supp{\Vec{z}(t)} = \supp{\Vec{p}(t)}$ for every $t \in [0, T]$.

Consider now a mesh $t_0 = 0< t_1 < \dots < t_N = T$ for the interval $[0, T]$ with step size $\tau = T/N$. 
Theorem~\ref{prop:change_of_variables} suggests the following recipe to obtain a sequence $(\Vec{z}^{(k)})_{k \in [N]}$ approximating the solution of \eqref{eq:ratio-ivp} on $[0, T]$. Define the sequences $(\Vec{p}^{(k)})_{k \in [N]}$ and $(\Vec{u}^{(k)})_{k \in [N]}$ via the recurrence
\begin{equation}\label{eq:pop-recurrence}
    \begin{cases}    
    \Vec{p}^{(0)} &=  \Vec{z}^{(0)}\\
    u_{i}^{(k)} &=  \deriv{f}{i}\left( (\sum_{j = 1}^n p_{j}^{(k)})^{-1} \Vec{p}^{(k)}\right), \quad i \in [n]\\
    p_{i}^{(k + 1)} &=  p_{i}^{(k)} \exp(\tau u_{i}^{(k)})
    \end{cases}
\end{equation}
and observe that $\vec{p}^{(k + 1)}$ is obtained from $\vec{p}^{(k)}$ by evaluating in $t_{k+1}$ the solution to the initial value problem
\begin{equation*}
    \begin{cases}
        \dot{y}_i &= y_i \deriv{f}{i}\left((\sum_{j = 1}^n p_j^{(k)})^{-1} \Vec{p}^{(k)}\right), \quad i \in [n]\\
        \Vec{y}(t_k) &= \Vec{p}^{(k)},
    \end{cases}
\end{equation*}
which is a linearization of \eqref{eq:pop-ivp} about $\Vec{p}^{(k)}$.
Now assuming $ \Vec{p} (t_k) \approx \Vec{p}^{(k)}$ and using that $\Vec{z}(t) = (\sum_{i = 1}^n p_i(t))^{-1} \Vec{p}(t)$
leads to the approximation $ \Vec{z} (t_k) \approx \Vec{z}^{(k)} = (\sum_{i = 1}^n p_i^{(k)})^{-1} \Vec{p}^{(k)}$. Substituting this expression in \eqref{eq:pop-recurrence} shows that $(\Vec{z}^{(k)})_{k \in [N]}$ satisfies the recurrence
\begin{equation}\label{eq:mwu}
    z_{i}^{(k + 1)}
        =  z_{i}^{(k)}
        \dfrac{\exp( \tau \deriv{f}{i}( \Vec{z}^{(k)})  )}
        {\sum_{j = 1}^n  z_{j}^{(k)}\exp( \tau \deriv{f}{j}( \Vec{z}^{(k)})  )}.
\end{equation}
The reader shall recognize in \eqref{eq:mwu}  one of the formulations of the multiplicative-weight-update rule \cite{Arora2012TheMW}.
Moreover,  this form of recurrence is well-known in the replicator dynamics framework \cite{weibull_evolutionary_1995}.

Observe that if $\Vec{z}^{(k)}$ is an element of $\Delta_n$ and the gradient of $f$ is defined in $\Vec{z}^{(k)}$, then $\Vec{z}^{(k+1)}$ is well defined and it is an element of $\Delta_n$, having the same support of $\Vec{z}^{(k)}$. In particular, if $\Vec{z}^{(0)} \in \relint{\Delta_n}$, then such recurrence can be extended indefinitely, and this requires exclusively that $f$ be differentiable in $\relint{\Delta_n}$. However, observe also that, in numerical implementations, the discrete dynamics may jump on $\partial \Delta_n$ due to floating point arithmetic.

%% file: sections/flow_for_i.tex
\section{Computing the capacity}\label{sec:contribution-i-specific}
\subsection{Optimization program for the capacity}\label{subsec:flow-for-i}
\noindent We now apply the framework described so far to tackle the computation of the capacity for the DMC.

Define $\Vec{z} = (z_1, \dots z_n) \in \Delta_n$ by
\begin{align*}
    z_i &= \prob{X}(x_i), &i \in [n]    
\end{align*}
and
$\vec{q} = \Vec{q}(\vec{z}) = (q_1, \dots q_m) \in \Delta_m$ by
\begin{align*}
    q_j &= \prob{Y}(y_j) = \sum_{i = 1}^{n} p(j|i) z_i, & j \in [m]
\end{align*}
so that $\vec{z}$ and $\vec{q}$ are the distribution of $X$ and $Y$ respectively.
In fact, $\prob{Y|X = x_i}(y_j) = p(j|i)$, and by the theorem of total probability
\begin{align*}
    \prob{Y}(y_j) &= \sum_{i = 1}^n p(j|i)\prob{X}(x_i), & j \in [m]
\end{align*}
Obviously $I[X;Y]$ can be rewritten as a function of $\Vec{z}$, since the distribution of $Y$ depends solely on the distribution of $X$ and on the transition matrix $\mat{P}$. In fact,  $I[X, Y] = I(\Vec{z})$, where 
\begin{equation}\label{eq:i-di-zeta}
    I(\Vec{z}) = \sum_{i = 1}^{n} c_i z_i - \sum_{j = 1}^{m} q_j \ln{q_j},
\end{equation}
being $c_i = \sum_{j = i}^{m} p(j|i) \ln[p(j|i)]$,
and the capacity $C$ is then the value of the program:
\begin{maxi}|s|[0]
    {\Vec{z} \in \Delta_{n}} 
    {I(\Vec{z})} 
    {\label{p:i-di-z}}
    {} 
\end{maxi}
Following \cite{boche_2023_10130754}, we call \emph{optimal input distribution} every global solution to \eqref{p:i-di-z}, \textit{i.e.}, every $\Vec{z} \in \Delta_n$ such that
\begin{equation*}
    I(\Vec{z})= C = \max_{\Vec{y} \in \Delta_n}{I(\Vec{y})}.
\end{equation*}

\subsection{Vector flow for capacity computation}\label{subsec:method-works}
\noindent As a consequence of well-known properties of the mutual information, the objective function $I(\Vec{z})$ is continuous and concave on $\Delta_n$.%
    \footnote{See Appendix~\ref{app:proofs} for a nice proof of this known result.}
Motivated by the results in Section~\ref{sec:general-flows}, we would like then to consider the ODE
\begin{align}\label{eq:ode-I-formal}
    \dot{z}_{i}
    &=
    z_{i}
    \left[ \deriv{I}{i}(\Vec{z}) - \sum_{k = 1}^n z_{k } \deriv{I}{k}(\Vec{z}) \right],
   & i\in [n].
\end{align}
as a means to address the DMC capacity computation. However, some care is required, for the results mentioned may be applied directly to \eqref{eq:ode-I-formal} only if $I(\Vec{z}) \in C^{1, 1}(\Delta_n)$. As shown in Proposition~\ref{prop:derivate}, the function $I(\Vec{z})$ can lack of this smoothness requirement, as it depends on the zero elements of transition matrix $\mat{P}$. Nevertheless, using some technical arguments, it can be shown that on the set on which the (forward) flow \eqref{eq:ode-I-formal} is well-defined, the (forward) dynamics is kept sufficently away from the set in which $I(\Vec{z})$ is not differentiable, and this allows to adapt the main ideas of Section~\ref{sec:general-flows} also to \eqref{p:i-di-z}.
\begin{theorem}[Attaining capacity]\label{thm:we-win}
    For every $\vec{y} \in \relint{\Delta_n}$ there exists a unique trajectory $\vec{z}(t)$ of \eqref{eq:ode-I-formal} satisfying $\Vec{z}(0)= \Vec{y}$, which converges to an optimal input distribution as $t \to +\infty$.
\end{theorem}
\begin{IEEEproof}
    For every $j \in [m]$, define
    \begin{equation*}
        \mathcal{S}_j = \set{i \in [n] | p(j|i) = 0}.
    \end{equation*}
    It is easy to see that for every $\tilde{\Vec{z}} \in \Delta_n$:
\begin{itemize}
    \item if $\mathcal{S}_j = \emptyset$, then $q_j(\tilde{\Vec{z}}) \geq \min_{i \in [n]}{p(j|i)} > 0$;
    \item if $\mathcal{S}_j \neq \emptyset$, then $q_j(\tilde{\Vec{z}})=0$ if and only if $\supp{\tilde{\Vec{z}}} \subseteq \mathcal{S}_j$. 
\end{itemize}
    In particular, if all the transition probabilities are positive, then this entails that $I(\vec{z})\in \mathcal{C}^{2}(\Omega)$ for some open $\Omega\subset \bbR^{n}$ that is a superset of $\Delta_n$, and Theorem~\ref{thm:concave} applies, since $I(\Vec{z})$ is concave.

    In the general case, consider the sets $V_{\epsilon} \subset \Omega_{\epsilon} \subset \bbR^{n}$ that depend on a parameter $\epsilon > 0$ and that are defined by
    \begin{align*}
        V_{\epsilon} &= \set{\Vec{z} \in \bbR^n | q_j(\Vec{z}) \geq  2\epsilon, \; j \in [m]} \cap [ 0,  +\infty)^{n}\\
        \Omega_{\epsilon} &= \set{\Vec{z} \in \bbR^n | q_j(\Vec{z}) > \epsilon, \; j \in [m]} \cap (-\epsilon, +\infty)^{n}.
    \end{align*}
    By construction, $V_{\epsilon}$ is a closed subset of $\bbR^{n}$, and $\Omega_{\epsilon}$ is an open subset of $\bbR^{n}$.
    Notice that $q_j(\Vec{y}) > 0$ for every $j$, hence $\Vec{y} \in V_{\epsilon}$ provided $\epsilon$ is sufficiently small.
    Observe that $I(\Vec{z}) \in \mathcal{C}^{1,1}(\Omega_{\epsilon})$, and so the Picard-Lindel\"{o}f's theorem is applicable to the ODE \eqref{eq:ode-I-formal} on $\Omega_{\epsilon}$. Therefore, there exists a maximum $T \in (0, +\infty]$ such that the initial value problem
    \begin{equation}\label{eq:ivp-I}
    \begin{cases}
        \dot{z}_i &= z_{i} \left[ \deriv{I}{i}(\Vec{z}) - \sum_{j = 1}^n z_j \deriv{I}{j}(\Vec{z})\right], \quad i \in [n]\\
        \Vec{z}(0) &= \Vec{y}
    \end{cases}
    \end{equation}
    admits one and only one solution $\Vec{z}(t) \in \Omega_{\epsilon}$ that is defined for $t \in [0, T)$.
    Moreover, the regularity condition ensures that $\Vec{z}(t) \in \Delta_n$ for  $t \in [0, T)$.

    Set $q_j(t) = q_j(\Vec{z}(t))$ for every $t \in [0, T)$. The idea, now, is proving that $q_j(t)\geq 2\epsilon$ for every $t \in [0, T)$ and every $j$ provided $\epsilon$ is sufficiently small. To this end, fix an arbitrary $j \in [m]$.
    Since $-\ln{u} \to +\infty$ as $u \to 0^{+}$, it is possible to find some $Q > 0$, possibly very small, such that for every $i \in \mathcal{S}_j$ 
    \begin{equation}\label{eq:q-bound}
        c_i - 1 - p(j|i)\ln{Q} - C > 0.
    \end{equation}
    Moreover, $q_j(t)> 0$ entails that $z_i(t) \neq 0$ for some $i \in \mathcal{S}_j$, thus using Proposition~\ref{prop:derivate},
    \begin{align*}
        &\dfrac{d}{dt}[q_j(\vec{z}(t))] 
        = \sum_{i = 1}^{n} \deriv{q_j}{i}(\vec{z}(t))\dot{z}_i(t)\\
        &\quad = \sum_{i = 1}^{n} p(j|i)
        z_i \left[c_i - 1 -\sum_{J = 1}^{m} p(J|i)\ln{q_J} - I(\Vec{z}) \right]\\
        &\quad = \sum_{i \in \mathcal{S}_j} p(j|i)
        z_i \left[c_i - 1 -\sum_{J = 1}^{m} p(J|i)\ln{q_J} - I(\Vec{z}) \right]\\
        &\quad \geq \sum_{i \in \mathcal{S}_j} p(j|i)
        z_i \left[c_i - 1 - p(j|i)\ln{q_j} - C \right],
    \end{align*}
    where the inequality $I(\Vec{z}) \leq C$ holds by definition of capacity, and $\ln{q_J}\leq 0$ for every $J$ since $0 < q_J \leq 1$.
    By \eqref{eq:q-bound}, for every $t \in [0, T)$, if $0 < q_j(t) \leq Q$, then $\dot{q}_j(t) > 0$. Therefore, let $\epsilon_j$ satisfying $0 < \epsilon_{j} < Q/2$, and observe that, as $t$ increases, then $\Vec{z}(t)$ cannot cross the hyperplane $H_{j}=\set{\Vec{z} \in \bbR^{n} | q_j(\Vec{z}) = 2 \epsilon_{j}}$.
    Consequently, $q_j(\Vec{z}(t)) > 2 \epsilon_{j}$ for every $t \in [0, T)$.
    
    This argument can be repeated for every $j \in [m]$, obtaining a corresponding $\epsilon_j$ as before. We then shrink $\epsilon$ so that it safisfies  $0 < \epsilon \leq \min_{j} \epsilon_j$. By construction, $\vec{z}(t)\in V_{\epsilon}$ for every $t \in [0, T)$. But then, $T = +\infty$, \textit{i.e.}, the solution to \eqref{eq:ivp-I} is defined on the interval $[0, +\infty)$. By Lemma~\ref{lem:mult}, it follows that $\vec{z}(t)$ admits a limit $\Vec{z}^{*} \in \Delta_n \cap \Omega_{\epsilon}$ and \eqref{eq:kkt-condition} holds, meaning that $\Vec{z}^{*}$ is a KKT point for \eqref{p:i-di-z}, in which the feasible set is convex and the objective function is concave by hypothesis. Therefore, $\Vec{z}^{*}$ is a global solution to \eqref{p:i-di-z}.
\end{IEEEproof}

\subsection{Connection with the Blahut-Arimoto algorithm}\label{sec:mwu-and-baa}
\noindent In 1972, Arimoto \cite{arimoto:capacity} and Blahut \cite{blahut:capacity} devised independently an iterative algorithm to compute the capacity of a DCT thanks to an alternating maximization procedure.
The Blahut-Arimoto algorithm (BAA) produces a sequence for the input distribution $\vec{z}^{(k)} \in \Delta_{n}$ and also a sequence $\vec{q}^{(k)} \in \Delta_{m}$ for the corresponding output distribution that is defined via $\vec{q}^{(k)} = \vec{q}(\vec{z}^{(k)})$. For every step $k$, the algorithm produces $\vec{z}^{(k+1)}$ by applying to $\vec{z}^{(k)}$ a map defined on $\PhiBA : \Delta_n \to \Delta_n$, here called \emph{Blahut-Arimoto map}.
What Arimoto and Blahut proved is that
\begin{equation*}
    \PhiBA^{\circ k}(\Vec{z}^{(0)}) = \underbrace{\PhiBA \circ \cdots \circ \PhiBA}_{\text{$k$ times}}(\Vec{z}^{(0)}) 
\end{equation*} converges to an optimal input distribution as $k \to +\infty$ if $\vec{z}^{(0)} \in \relint{\Delta_n}$.
We can finally prove a surprising relation linking the BAA with the flow that we have introduce to compute the capacity of a DMC. Theorem~\ref{thm:BAA} clarifies in what sense our approach can be regarded as a continuous version of the BAA.
\begin{theorem}\label{thm:BAA} 
    The Blahut-Arimoto map is a multiplicative-weight-update rule of the form \eqref{eq:mwu} where $\tau = 1$ and $f(\Vec{z}) = I(\vec{z})$ as in \eqref{eq:i-di-zeta}.
\end{theorem}
\begin{IEEEproof}
First, observe that
\begin{equation}\label{eq:deriv-i-di-zeta}
    \deriv{I}{i}(\Vec{z})
    = -1 + \sum_{j = 1}^m p(j|i) \ln \left(\frac{p(j|i)}{q_j(\Vec{z})}\right)
\end{equation}
as reported in Proposition~\ref{prop:derivate}.
The definitions occurring in step $k + 1$ of the BAA are the following \cite{arimoto:capacity}:
\begin{equation}\label{eq:arimoto}
    \begin{cases}
        m_{ij}^{(k + 1)} &= p(j|i) z_{i}^{(k)}/q_j^{(k)}\\
        r_i^{(k + 1)} &= \exp( \sum_{j = 1}^m  p(j | i) \ln m_{ij}^{(k+1)})\\
        S^{(k + 1)} &= \sum_{i = 1}^n r_i^{(k + 1)}\\
        C^{(k + 1)} &= \ln S^{(k + 1)}\\
        z_{i}^{(k + 1)} &= r_{i}^{(k + 1)} / S^{(k + 1)},\\
        q_j^{(k + 1)} &= \sum_{i = 1}^n p(j|i) z_{i}^{(k + 1)}.
    \end{cases}
\end{equation}
Starting from \eqref{eq:arimoto}, some elementary substitutions show that
\begin{equation*}
            S^{(k+1)} z_{i}^{(k + 1)} = \exp \left[ \sum_{j = 1}^m  p(j | i) 
            \ln \left( \dfrac{p(j|i) z_{i}^{(k)}}{q_j(\vec{z}^{(k)} )} \right) \right].
\end{equation*}
Then 
\begin{align*}
    \exp &\left[\sum_{j = 1}^m  p(j | i) \ln \left( \dfrac{p(j|i) z_i^{(k)}}{q_j(\vec{z}^{(k)} )} \right) \right]\\
    =&
    \exp \left[ \sum_{j = 1}^m  p(j | i) \ln \left( \dfrac{p(j|i) }{q_j(\vec{z}^{(k)} )} \right)
    + \ln(z_{i}^{(k)}) \right]\\
    =&
    z_{i}^{(k)} \exp \left[ \sum_{j = 1}^m  p(j | i) \ln \left( \dfrac{p(j|i) }{q_j(\vec{z}^{(k)} )} 
    \right) \right]
\end{align*}
and using also \eqref{eq:deriv-i-di-zeta} we get:
\begin{equation*}
    S^{(k+1)} z_{i}^{(k + 1)} = z_{i}^{(k)} \exp \left[\deriv{I}{i}(\Vec{z}^{(k)}) + 1 \right]
\end{equation*}
that can be rewritten as:
\begin{equation*}
    C z_{i}^{(k + 1)} = z_{i}^{(k)} \exp \left[\deriv{I}{i}(\Vec{z}^{(k)})\right]
\end{equation*}
by setting $C = S^{(k+1)} e^{-1}$. As $z_{i}^{(k + 1)} \in \Delta_n$, this means that
\begin{equation*}
    z_{i}^{(k + 1)}
        =  z_{i}^{(k)}
        \dfrac{\exp( \deriv{I}{i}( \Vec{z}^{(k)})  )}
        {\sum_{j = 1}^n  z_{j}^{(k)}\exp( \deriv{I}{j}( \Vec{z}^{(k)})  )}.
\end{equation*}
\end{IEEEproof}
\input{sections/circuit}

%% file: sections/circuit.tex
\subsection{Analog implementation}
\begin{figure}[ht]
    \centering
\includegraphics[width=\linewidth]{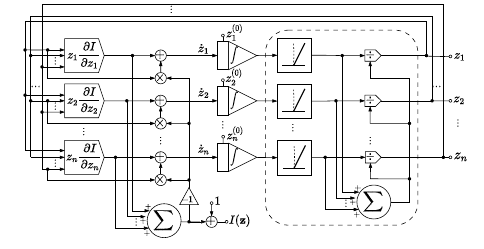}
    \caption{Circuit with normalizing module.}
    \label{fig:circuit}
\end{figure}

\noindent An attractive advantage of continuous-time methods is their amenability to be mapped onto hardware circuits~\cite{hopfield_neurons_1984,Cichocki1993Neural}. Figure~\ref{fig:circuit} illustrates a high-level implementation of the circuit logic associated with the procedure presented to compute the capacity of a DMC, given its transition matrix $\mat{P}$.

The signals $z_1(0)$, $\dots$, $z_n(0)$ are required as input, and represent a point in $\relint{\Delta_n}$ in which a trajectory of \eqref{eq:ode-I-formal} is initialized. Moreover, the circuit requires, for every $k \in [n]$, a module to implement $\Vec{z} \mapsto z_k\deriv{I}{k}(\Vec{z})$, that we here consider as a black-box, and that depends on  $\mat{P}$ as described in Proposition~\ref{prop:derivate}.
The signals $z_1$, $\dots$, $z_n$, which at time $t = 0$ coincide with $z_1(0)$, $\dots$, $z_n(0)$, evolve as time runs according to \eqref{eq:ode-I-formal} and can also be measured as output. The evolution of the signals is determined by the recurrent design of the circuit, that at time $t$ outputs the signals $z_1, \dots, z_n$ satisfying Volterra's equation \cite{birkhoff_rota_ODE}:
\begin{equation*}
    z_i = z_i(0) + \int_{0}^{t} \!z_i \deriv{I}{i}(z_1, \dots z_n)  - z_i \sum_{j = 1}^n z_j \deriv{I}{j}(z_1, \dots z_n) \,dt.
\end{equation*}
Such behavior is possible thanks to the integrator elements appearing in the circuit \cite{Cichocki1993Neural}.
The previous results guarantee that the output signals converge to the limit of the trajectory initialized in $(z_1(0), \dots, z_n(0))$, \textit{i.e.}, to an optimal input distribution.
At the same time, the circuit computes $\sum_{j = 1}^n z_j \deriv{I}{j}(z_1, \dots z_n) + 1$, which equals $I(\Vec{z})$ --- this follows by Proposition~\ref{prop:derivate} contained in Appendix~\ref{app:proofs}.
Therefore, it is possible to monitor the value of the objective function, which converges to the capacity of the DMC.
\begin{figure}[ht]
    \centering
    \includegraphics[width=\linewidth]{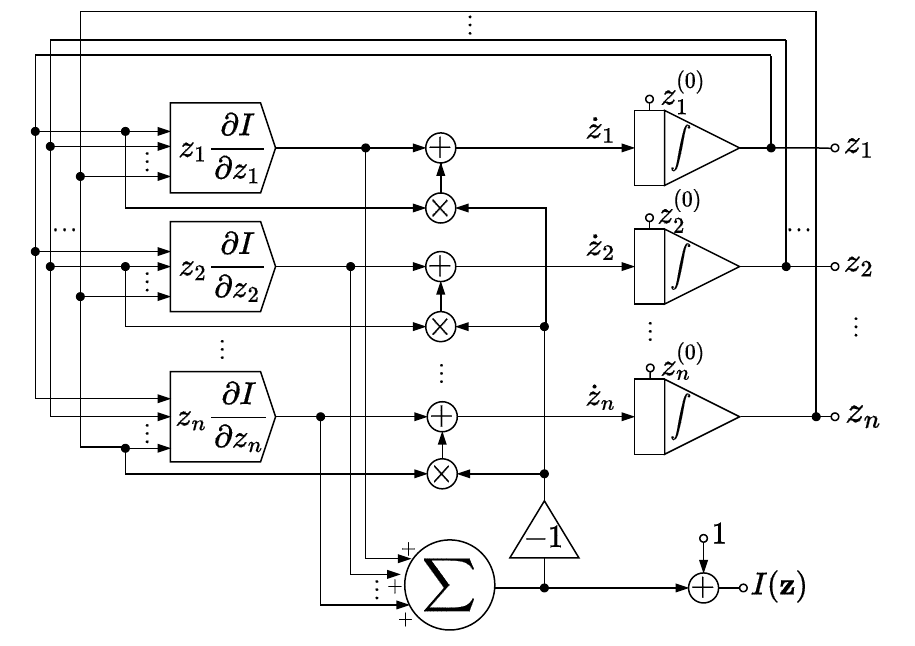}
    \caption{Ideal circuit.}
    \label{fig:circuit-wo-norm}
\end{figure}
Some remarks are necessary for actual implementation of the proposed circuit. Besides the black-box modules described before, another aspect affecting the applications is the stability of the underlying dynamical system. 
Theoretically, the analog implementation of the continuous-time dynamics remains within $\Delta_n$. However, real circuits may introduce external noise, potentially causing the signal to deviate from this constraint, and the theoretical results that we have presented neglect the stability of the ODE considered in a neighborhood of $\Delta_n$, which may be a critical factor if a perturbation causes the dynamics to leave $\Delta_n$. 
To mitigate this issue in case such deviations occur, we introduce in Fig~\ref{fig:circuit} a variation of the circuit considered before. This circuit is provided with a normalization module that mimics what we have carried out in numerical experiments, as discussed in Section \ref{sec:experiments}.
On each signal that the normalization module receives as input, a Rectified Linear Unit (ReLU) function, which sets any negative signals to zero, is applied. Subsequently, the resulting vector of non-negative signals is normalized with respect to the $L^1$-norm.

%% file: sections/experiments.tex
\section{Experiments}
\noindent 
Section~\ref{subsec:method-works} shows theoretical results for which the flow of \eqref{eq:ode-I-formal} converges towards an optimal input distribution. If this holds, then evaluating $I(\vec{z})$ along the trajectory leads, in the limit, to the channel capacity.
In this complementary section, we delineate the experiments that we have carried out, aiming to investigate the numerical implementation of the proposed vector flow optimization method. What follows is a comprehensive analysis to assess the effectiveness of employing an ODE solver on \eqref{eq:ode-I-formal} for estimating the capacity of a DMC by following the flow.
\smallskip 

\myparagraph{Dataset} 
We have generated a dataset containing (transition matrices of) \emph{symmetric} channels \cite{cover_elements_2006}. Symmetric channels are a particular class of DMC in which input and output alphabets coincide, and such that, in the transition matrix, every row is a permutation of every other row, and every column is a permutation of every other column.
For these channels, the uniform distribution on the input alphabet is an optimal input distribution, hence the capacity is readily computed by evaluating $I(\Vec{z})$ in the barycenter of $\Delta_n$.

Each transition matrix $\mat{P}$ in the dataset has been generated by perturbing the noiseless channel, for which the transition matrix is the identity matrix. The matrices generated depend on the \emph{dimensionality} $n (= m)$ in a set of $15$ evenly spaced integers within the interval $[2, 100]$, and on a parameter $\sigma \in \set{0, 0.25, 0.5, 0.75, 1}$ controlling the level of noise. For each value of $n$ and $\sigma$, we generated $15$ random symmetric channels $\mat{R}_n$ with dimensionality $n$ and added to the dataset the transition matrix
\begin{equation*}
    \mathbf{P} = (1 - \sigma) \mat{I}_n + \sigma \mat{R}_n.
\end{equation*}
The generation of every matrix $\mat{R}_n$ has been performed by first using a Dirchlet distribution \cite{ng_dirichlet_2011} to generate the first row of $\mat{R}_n$ uniformly in $\Delta_n$, and then by permuting cyclically the row generated to obtain the remaining rows of $\mat{R}_n$.

\subsection{Vector Flow Optimization Algorithm}
\subsubsection{ODE solver} The trajectories of \eqref{eq:ode-I-formal} have been discretized using an ODE solver that implements a modified version of the Euler method.
When operating on an ODE of $n$ variables, the solver alternates a step of the Euler method with step size $\tau$, which is given as input to the solver, and a normalizing step, that ensures that the dynamics stays in $\Delta_n$. The following are the two operations performed during the normalizing step: 
\begin{enumerate}
    \item first, the ReLU function is applied to every entry of the vector $\vec{z}$ produced by the Euler method, meaning that $z_i$ is replaced with $z_i^{+} = \max\set{z_i, 0}$;
    \item then, the resulting vector is normalized with respect to the $L^1$-norm.
\end{enumerate}
The solver, for a given ODE on $n$ variables and some $\Vec{z}^{(0)} \in \Delta_ n$, produces the corresponding discretized trajectory initialized in $\Vec{z}^{(0)}$ by applying iteratively the adjusted Euler method with stopping condition $\norm{\Vec{z}^{(k)} - \Vec{z}^{(k+1)}}_1 < 10^{-4}$, where $\Vec{z}^{(k+1)}$ is obtained from $\Vec{z}^{(k)}$ at iteration $k$, and in any case for maximum $10\,000$ iterations.
For a more in-depth elucidation on the adjusted Euler method, we direct the reader to consult Appendix~\ref{app:odeNorm}, which explores the applications of the adjusted Euler method within the context of our study.\smallskip

\subsubsection{Interior-point algorithm}
We now describe the algorithm that we have utilized to compute the capacity of a DMC. The algorithm, which we shall call vector-flow algorithm in the sequel, requires as input the transition matrix of a DMC. It then approximates the corresponding ODE \eqref{eq:ode-I-formal} using the solver described above, and where the initialization point $\Vec{z}^{(0)}$ is chosen uniformly in $\relint{\Delta_n}$ via a Dirichlet distribution, being $n$ the dimensionality of the channel.
Clearly, as the loop terminates, $k$ equals the number of iterations actually performed during the loop, and 
then $\hat{C} = I(\Vec{z}^{(k)})$ is the estimation for the capacity that the vector-flow algorithm outputs.
\subsection{Experimental setup}\label{sec:experiments}
\noindent 
\subsubsection{Investigation on the step size}
Firstly, We have carried out an investigation on the step size $\tau$ using the generated DMC dataset.
The objective of this first phase was to understand how the value of $\tau$ affects 
the optimization task. To this end, we have utilized $25$ evenly spaced values of $\tau$ in the interval $[0.01, 30]$. For each value of $\tau$ and each channel, we have estimated the capacity with the vector-flow algorithm and compared it against the ground truth. Moreover, to investigate the speed of convergence, we have studied also how the number of iterations varies for different values of $\tau$. As a benchmark, we launched the BAA, using the same stopping criterion adopted in the vector-flow algorithm, and counted the number of iterations executed. To have a fair comparison, both algorithms trajectories have been initialized in the same point of $\relint{\Delta_n}$.
\smallskip

\subsubsection{Comparative analysis}
After that, we selected $\tau = 1$, a choice that is theoretically sound by Theorem~\ref{thm:BAA},  and inspected how the objective function behaves at every iteration.
For the DMCs considered, we have compared again the vector-flow algorithm capacity estimation against the ground truth. We then compared the performance of the vector-flow algorithm against the BAA. As in the investigation on the step size, the same initialization and the same stopping criterion has been used for both algorithms.

\smallskip

\subsubsection{Trajectory plots}
We also wanted to explore visually how the dynamics of the proposed method differs from another constrained optimization method, the projected gradient ascent \cite{luenberger_linear_2016}. To this end, we considered some  trajectories evolving on $\Delta_3$. Two DMCs have been considered for the plots:
\begin{itemize}
    \item the symmetric channel with transition matrix
\begin{equation}
    \begin{bmatrix}
    0.3668 & 0.5678 & 0.0654\\
    0.5678 & 0.0654 & 0.3668\\
    0.0654 & 0.3668 & 0.5678
    \end{bmatrix}, \notag
\end{equation}
for which the barycenter of $\Delta_n$ is the unique optimal input distribution, as well as the corresponding output distribution;
\item the ternary confusion channel \cite{mackay_information_2003}, for which $n = 3$ and $m = 2$, and which has transition matrix
\begin{equation}
    \begin{bmatrix}
        1 & 0\\
        0 & 1 \\
        0.5 & 0.5
    \end{bmatrix}.\notag
\end{equation}
The vector $(0.5, 0.5, 0)$ is the unique optimal distribution for the ternary confusion channel, and $(0.5, 0.5)$ is the corresponding output distribution.
\end{itemize}

%% file: sections/results.tex
\subsection{Experimental Results}\label{sec:results}

\subsubsection{Investigation on the step size}
\begin{figure*}[ht!]
    \centering
    \subfloat[]{
        \includegraphics[width=0.48\linewidth]{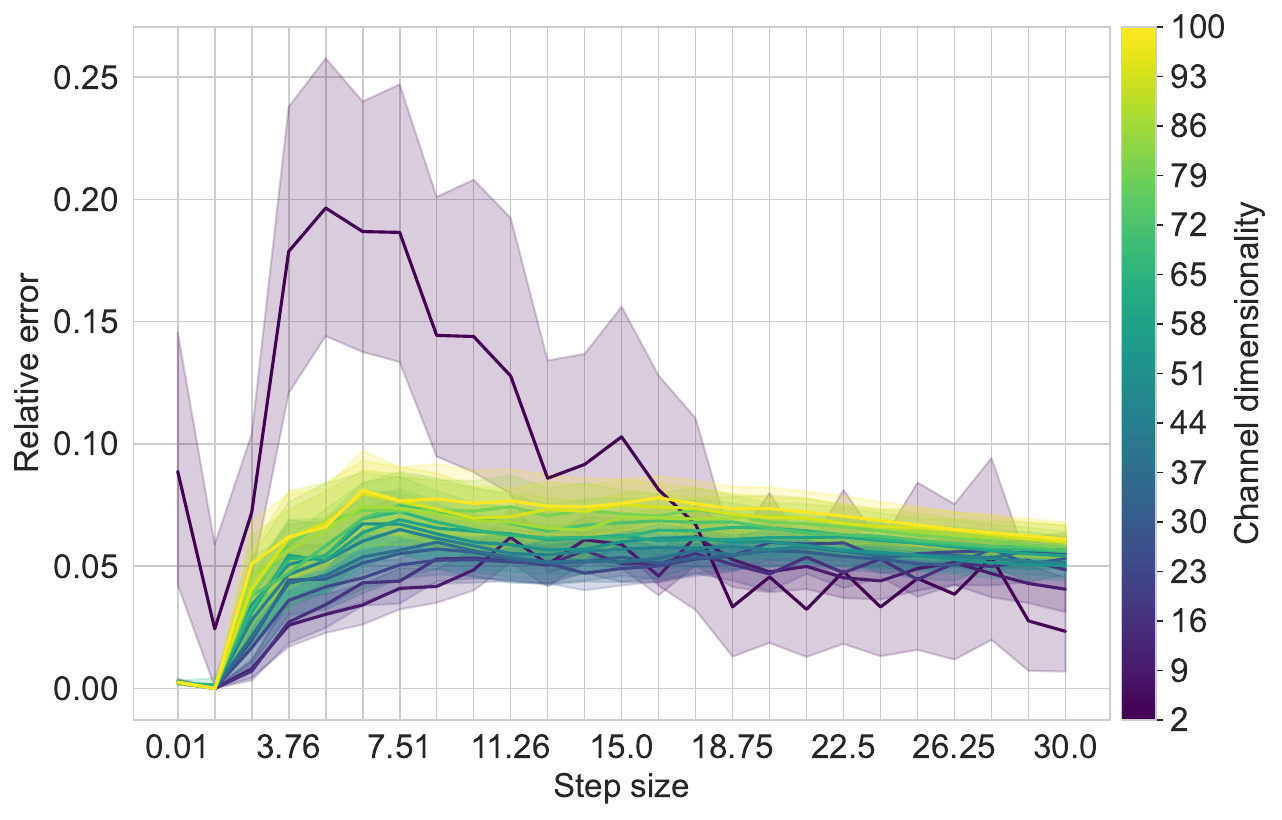}\label{fig:err}
    }
    \hfill
    \subfloat[]{
    \includegraphics[width=0.48\linewidth]{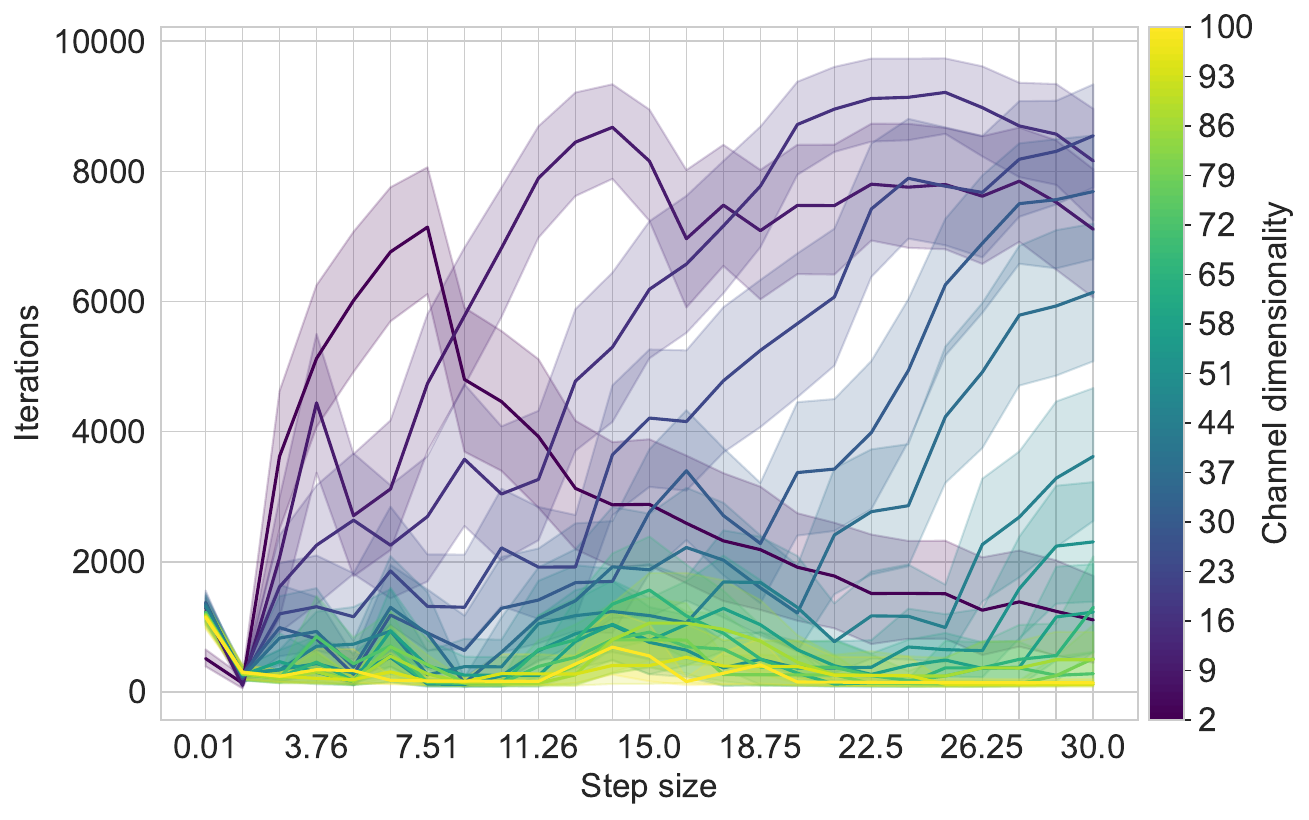}\label{fig:steps}
    }
    \caption{Investigation on the step size. \protect\subref{fig:err} Relative error of the capacity estimation against the ground truth; \protect\subref{fig:steps} Number of iterations executed.}
    \label{fig:stepsdiff}
\end{figure*}

Figure~\ref{fig:err} shows the relative error of the capacity estimation against the ground truth for different values of $\tau$.
The plots show that for $\tau$ close to $1$, the curves present a minimum reaching a height close to $0$. This means that for $\tau$ close to $1$ the estimation is very precise. Besides that, the figure shows that for almost every dimensionality the relative error is on average below $0.1$, with the exception of $n = 2$, for which the peak of relative error exceeds $0.25$ and then the relative error drops below $0.1$ as $\tau$ increases.

Figure~\ref{fig:steps} represents the number of iterations executed for different values of $\tau$. The plots present an oscillatory behavior that occurs with smaller amplitude as the dimensionality increases.
What can be seen is that for values $\tau$ close to $1$, the number of iterations remains in the order of $10^2$ for every value of $n$.

\begin{figure}[ht!]
    \centering
    \includegraphics[width=\linewidth]{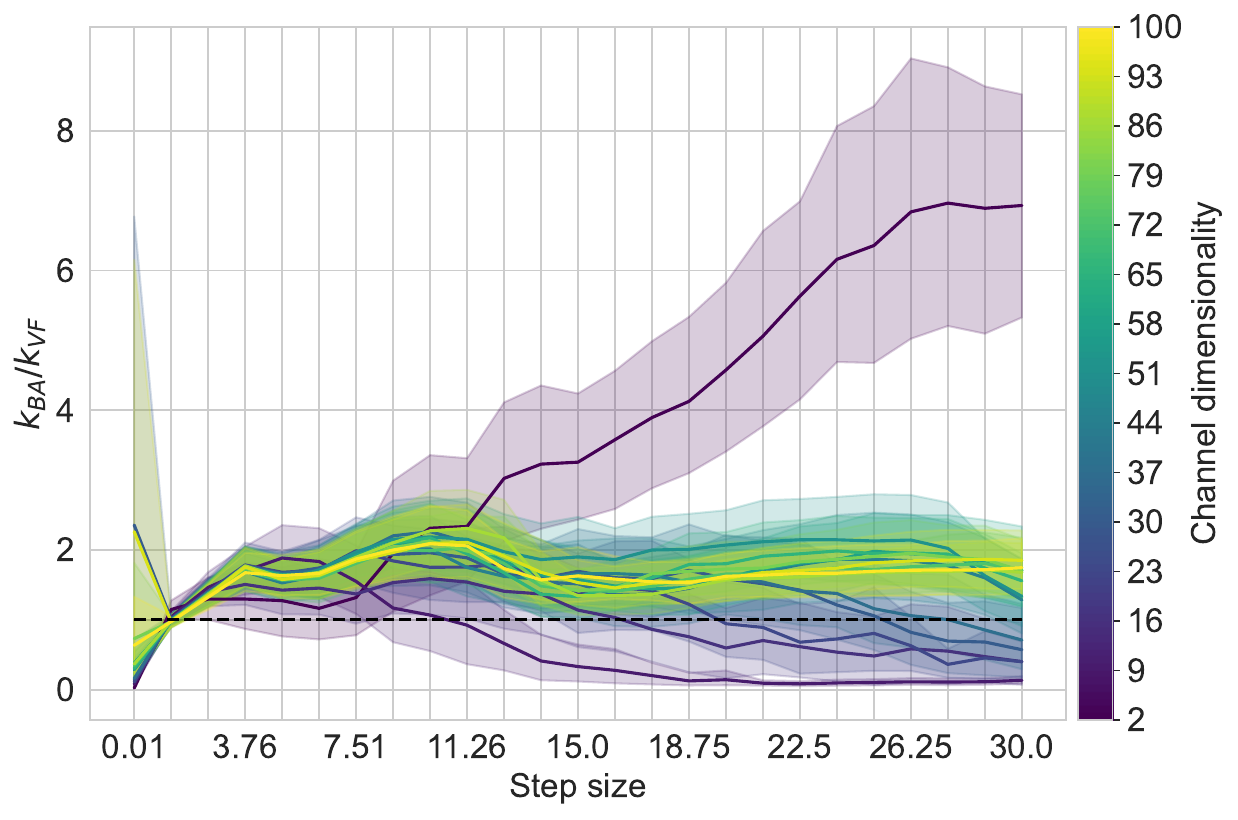}
    \caption{Ratio of the number of iterations executed with the BAA and with the vector-flow algorithm.}
    \label{fig:stepsdiffratio}
\end{figure}

We also compared the number of iterations against those of the BAA. As shown in Figure \ref{fig:stepsdiffratio}, the ratio of the number of iterations indicates that the vector-flow algorithm takes, generally speaking, no more than three times the number of iterations of the BAA, with the exception of $n = 2$ and for the minimum value of $tau$ considered. \smallskip

\subsubsection{Comparative analysis}
\begin{figure}[ht!]
     \centering
     \includegraphics[width=\linewidth]{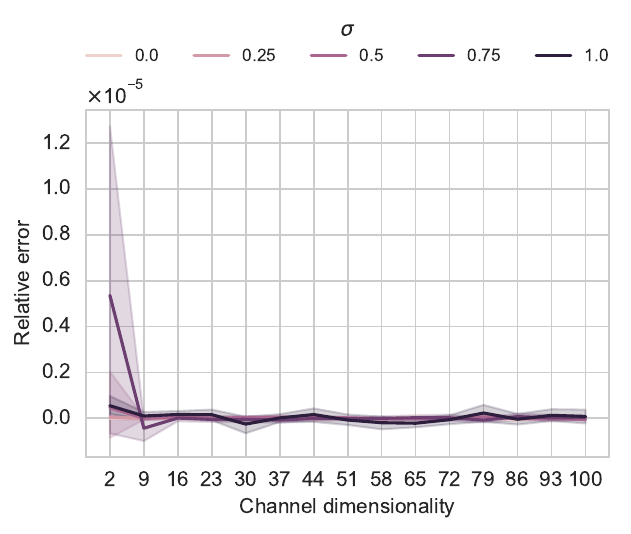}
     \caption{Relative error of the capacity estimation against the ground truth for $\tau = 1$}  \label{fig:random-channel-capacity}
\end{figure}

\noindent For every random DMC and step size considered, we have compared the capacity estimation obtained with the vector-flow algorithm for $\tau=1$ and the actual value. As illustrated in Figure~\ref{fig:random-channel-capacity}, the relative error of the capacity estimation does not exceed $1.5 \times 10^{-5}$, and for $n > 9$ it stays in the order of $10^{-7}$. The error is so small that sometimes the value obtained is even larger than the theoretical result computed. This paradoxical outcome probably hides some approximation errors that occur during the computation of the objective function and deserves further investigation. Besides that, this shows that for $\tau = 1$ the capacity estimation is extremely close to the ground truth in the instances considered.

Comparing the number of iterations against the BAA, what emerges is that the two algorithms take essentially the same number of iterations on noisy channels. For these DMCs, the vector-flow algorithm requires up to $20\%$ more iterations than the BAA for $n = 2$, whereas for $n \geq 9$ the BAA takes up to $10\%$ iterations less than the BAA.

In contrast, on noiseless channels, the vector-flow algorithm outperforms the BAA, requiring fewer iterations to stop, and the gap increases with the dimensionality. In fact, by plotting the ratio of the number of iterations against the number of iterations of the BAA, the resulting graph resembles exponential decay.
This demonstration highlights the effectiveness of the procedure in solving the desired optimization task.

\begin{figure}[ht!]
    \centering
    \includegraphics[width=\linewidth]{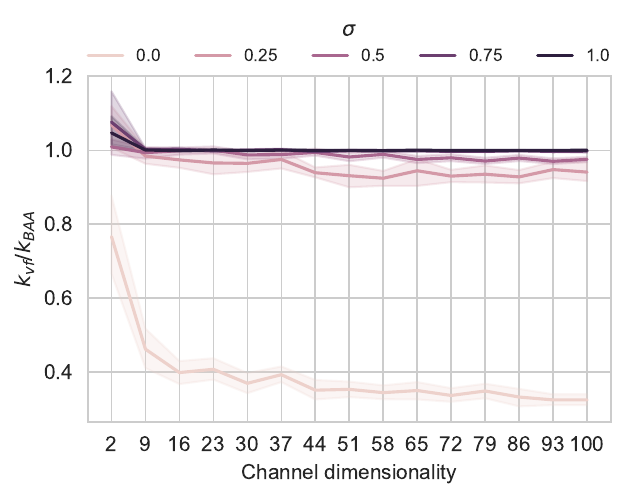}
    \caption{Ratio of number of iterations: vector flow algorithm for $\tau= 1$ against BAA.}    
    \label{fig:random-channel-steps}
\end{figure}

\subsubsection{Trajectory plots}
\begin{figure}[ht!]
    \centering
    \includegraphics[width=\linewidth]{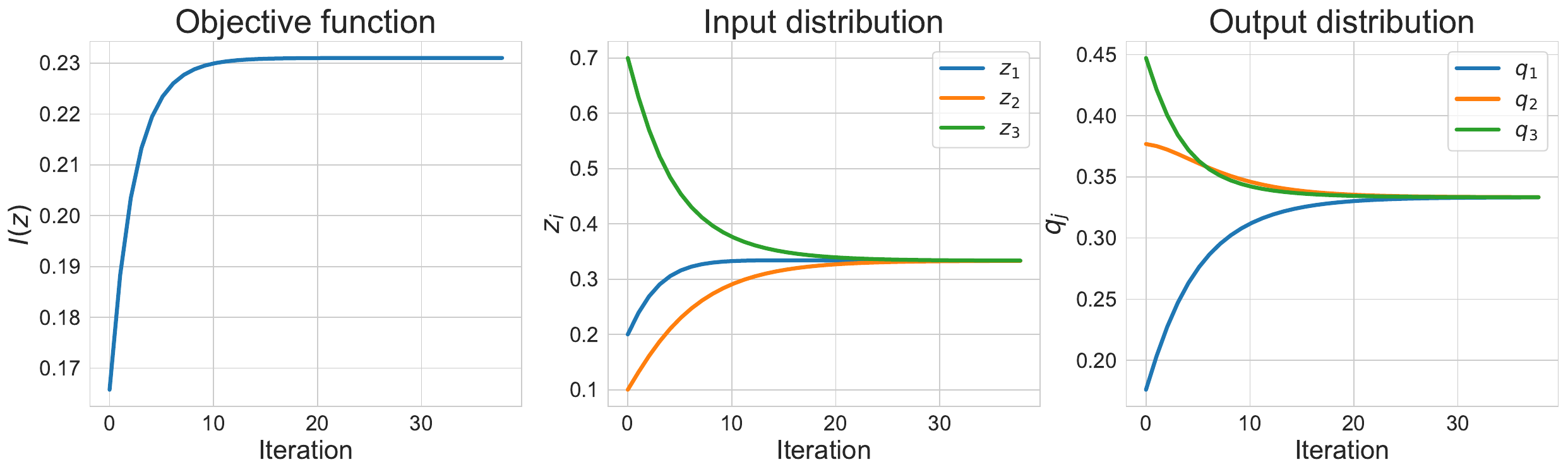}
    \caption{Estimating the capacity for the symmetric channel. The plots show the evolution of the input distribution, the corresponding output distribution, and the associated value of the objective function.}
    \label{fig:symmetric_channel}
\end{figure}

\begin{figure}[ht!]
    \centering
    \subfloat[]{
        \includegraphics[width=0.45\linewidth]{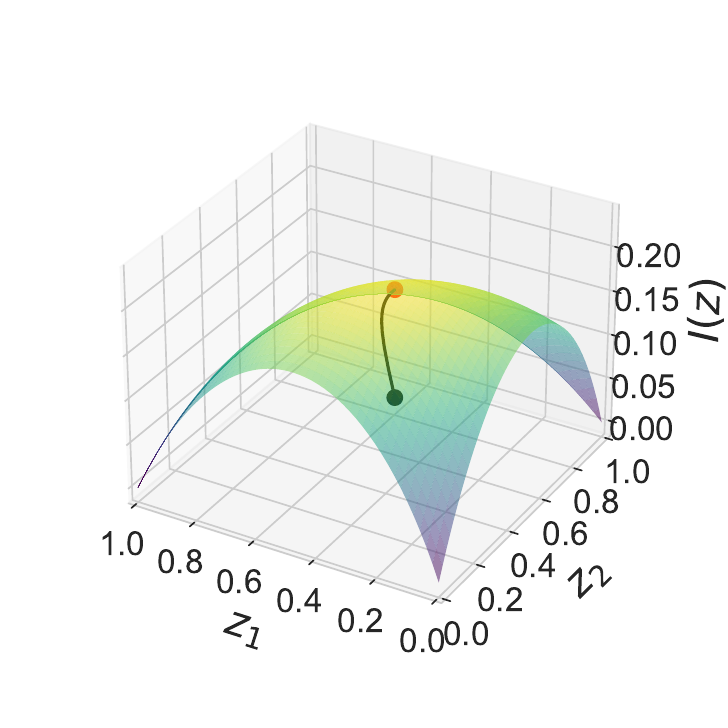}\label{fig:sym_3d}
    }
    \hfill
    \subfloat[]{
        \includegraphics[width=0.45\linewidth]{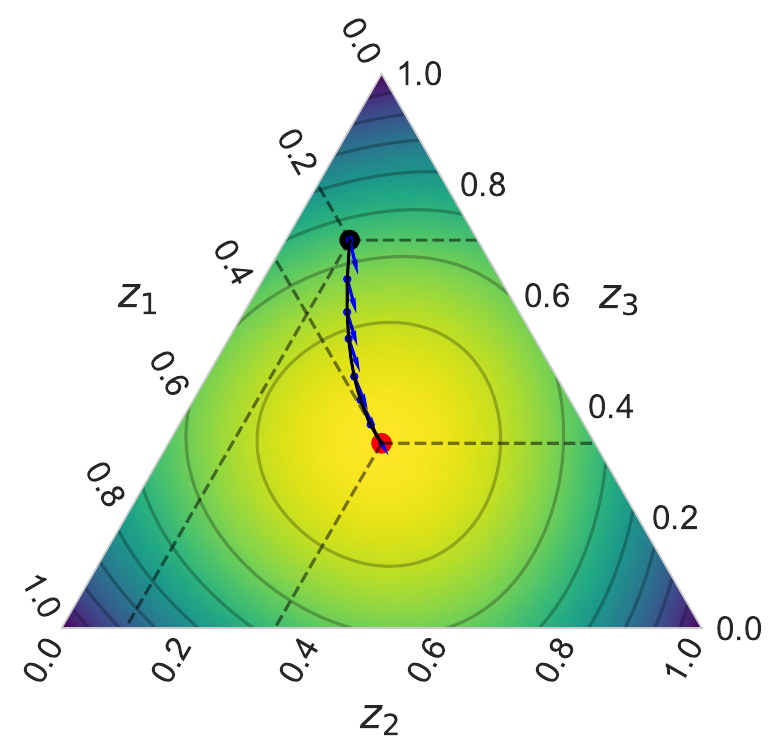}\label{fig:sym_proj}
    }
    \caption{Optimization trajectory for the capacity of the symmetric channel: \protect\subref{fig:sym_3d} 3D representation; \protect\subref{fig:sym_proj} 2D representation.}
    \label{fig:symmetric_channel_trajectory}
\end{figure}

\noindent Figure~\ref{fig:symmetric_channel} and Figure~\ref{fig:symmetric_channel_trajectory} describe the experiments involving the symmetric channel on $3$ symbols considered.
Figure~\ref{fig:symmetric_channel} illustrates that the value of the objective function increases strictly along the discrete trajectory and that the method is effective in computing the capacity of the channel. It is also apparent that the trajectory converges to the optimal input distribution.
\begin{figure}[ht!]
    \centering
    \includegraphics[width=\linewidth]{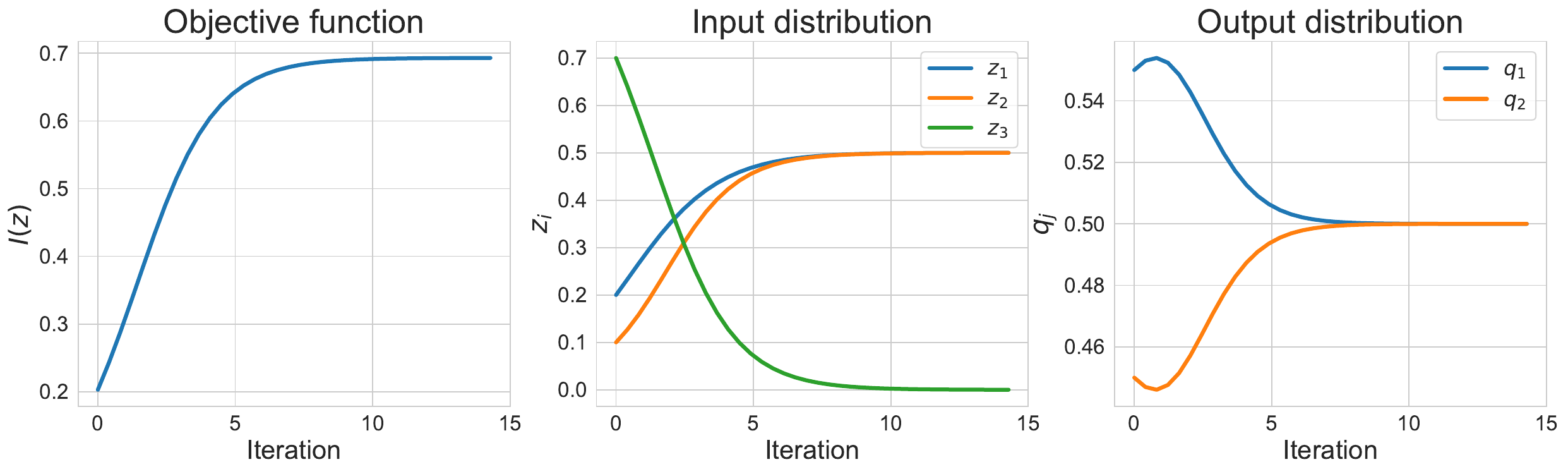}
    \caption{Estimating the capacity for the ternary confusion channel. The plots show the evolution of the input distribution, the corresponding output distribution, and the associated value of the objective function.}
    \label{fig:ternary_channel}
\end{figure}
A 3D representation and a 2D representation of the trajectory are displayed in Figure~\ref{fig:symmetric_channel_trajectory}. In both representations, the starting point is denoted with a dot, the final point with a red marker, and the colors represent the value of the objective function. The 3D representation enables to visualize how $(z_1,z_2, I(\Vec{z}))$ evolves along the dynamics. 
The 2D representation portrays an isometrical 2D projection of $\Delta_3$ and how the trajectory evolves with respect to the level curves of the objective function. 
This plot includes also the direction of the gradient of $I(\Vec{z})$ projected on the plane $\set{z_1 + z_2 + z_3 = 1}$, illustrating that the projected gradient is not parallel to the vector field defining the dynamics. It is also apparent that the cosine of the angle they form is positive along the trajectory. This behavior aligns with Theorem~\ref{thm:lyapunov}.

\begin{figure}[ht!]
    \centering
    \subfloat[]{
        \includegraphics[width=0.45\linewidth]{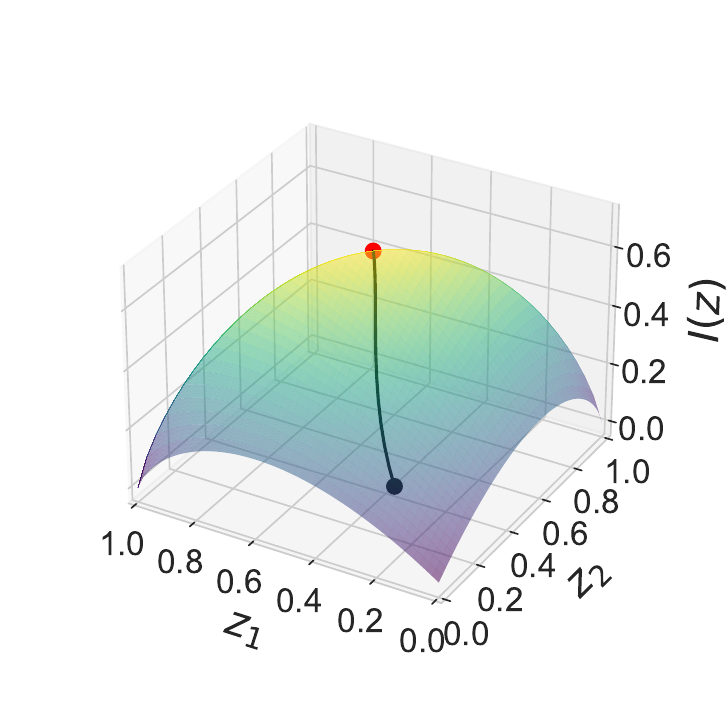}\label{fig:tern_3d}
    }
    \hfill
    \subfloat[]{
        \includegraphics[width=0.45\linewidth]{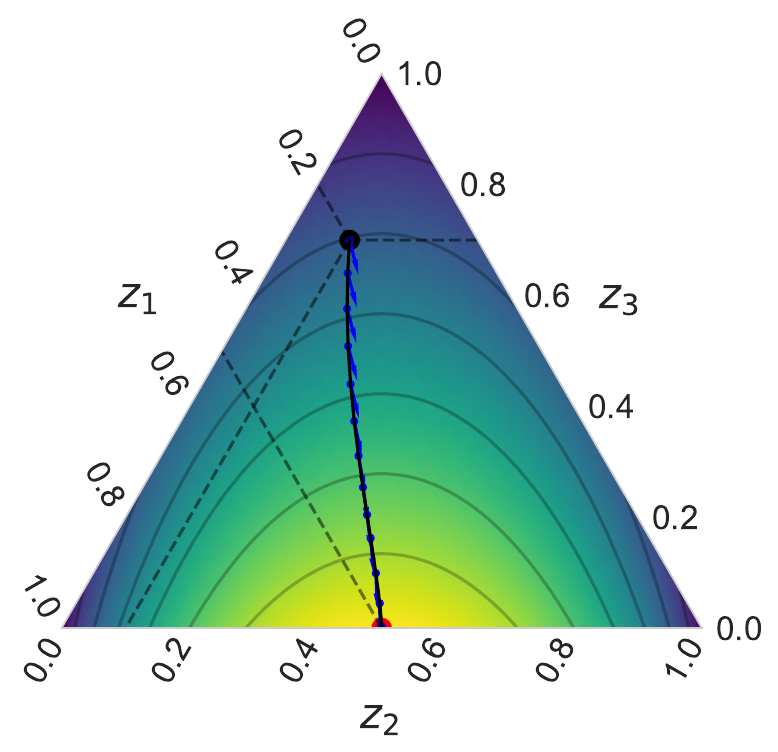}\label{fig:tern_proj}
    }
    \caption{Optimization trajectory for the ternary confusion channel:\protect\subref{fig:tern_3d} 3D representation; \protect\subref{fig:tern_proj} 2D representation.}
    \label{fig:ternary_channel_trajectory}
\end{figure}

Similar conclusions are drawn in the case of the ternary symmetric channel, as displayed in Figure~\ref{fig:ternary_channel} and Figure~\ref{fig:ternary_channel_trajectory}.

%% file: sections/discussion.tex
\section{Discussion}\label{sec:discussion}
\noindent That the Blahut-Arimoto map $\PhiBA$ acts as a multiplicative-weight-update rule is a known fact (see \textit{e.g.} \cite{Matz_2004_1405276,Naja2009Geometrical}). However, we are not aware of any work in which $\PhiBA$ is derived by discretizing a vector flow associated with the function $I(\vec{z})$.

The experiments conducted demonstrate that the algorithm presented is an effective discretization of the vector flow introduced in Section~\ref{subsec:flow-for-i} as far as the capacity computation is concerned. With a suitable choice of $\tau$, the method can reach an estimated value for the capacity that is comparable to the BAA benchmark, often requiring less iterations.
The plots of Figure~\ref{fig:random-channel-capacity} represent an additional confirmation that our method, using a suitable value for $\tau$, is essentially equivalent to the BAA, and this demonstrates that the vector-flow algorithm is effectively solving the desired optimization task.
The fact that the capacity estimated with the BAA matches with the estimation given by $\tau = 1$ demonstrates the quality of the approximation described in Theorem~\ref{thm:BAA}.

%% file: sections/conclusion.tex
\section{Conclusion}\label{sec:conclusion}
\noindent This paper discusses an approach to address the capacity computation of a DMC by means of a suitably defined vector flow. Proofs are provided to demonstrate that the introduced continuous-time dynamics converge to an optimal input distribution.
A possible scheme of a circuit has been discussed enabling to perform analog computations to estimate the capacity.
It is also explained how an analogy with replicator dynamics can be used to claim that the introduced flow can be regarded as a continuous-time version of the BAA. Furthermore, the conducted experiments demonstrate that a suitable adjustement of the Euler method allows to simulate the flow previously discussed and that can effectively estimate the capacity of DMCs

Among possible future directions of research, the application of different numerical methods to the ODE presented may be interesting, as it may be used to produce new algorithms for estimating the capacity of the DMC, as well as for estimating an optimal input distribution.
It may also be interesting to consider DMC subject to some constraints, as in \cite{blahut:capacity}, and see to what extent the flow here introduced can be adapted to that more challenging setting. 

It may also be worth investigating some more advanced techniques to improve the circuit presented, which could hopefully result in actual implementation of the idea presented.

%% file: sections/proofs.tex
\noindent This appendix reports some proofs, computations, and technical results that have been omitted in the previous sections for the sake of the exposition.
\begin{proposition}\label{prop:kkt-equivalence}
    Let $f$ be a continuous function defined on $\Delta_{n}$ and suppose $f$ is differentiable in a neighborhood $\Omega$ of $\Vec{y} \in \Delta_n$. Then $\Vec{y}$ satisfies the KKT condition for the program
    \begin{maxi}|l|[0]
        {\Vec{z} \in \Delta_{n}} 
    {f(\Vec{z})} 
    {\notag}
    {} 
    \end{maxi}
    if and only if there exists some $C \in \bbR$ such that
    \begin{equation}\label{eq:kkt-condition}
        \deriv{f}{i}(\vec{y})
    	\begin{cases}
            = C, & y_{i} > 0\\
            \leq C, & y_{i} = 0.
        \end{cases} 
    \end{equation}
\end{proposition}
\begin{IEEEproof}
    The Lagrangian associated with the maximization program is the function
    \begin{equation*}
        \mathcal{L}(\Vec{z}, \Vec{a}, b) = f(\Vec{z}) + \sum_{i = 1}^n z_{i} a_{i} + b\left( 1- \sum_{i = 1}^n z_{i}\right)
    \end{equation*}
    where $(\Vec{z}, \Vec{a}, b) \in \Omega \times \bbR_{+}^n \times \bbR$.
    The KKT condition \cite{luenberger_linear_2016} leads to the system of equations
    \begin{equation}\label{eq:kkt-sys}
        \begin{cases}
        \dfrac{\partial \mathcal{L}}{\partial a_i}(\Vec{z}, \Vec{a}, b) &= \deriv{f}{i}(\Vec{z}) + a_i - b = 0\\
        \sum_{i = 1}^n z_i a_i &= 0\\
        \Vec{z} &\geq \Vec{0}\\
        \Vec{a} &\geq \Vec{0}\\
        \sum_{i = 1}^n z_{i} &= 1.\\
        \end{cases}
    \end{equation}
    Observe now that for $\Vec{z} = \Vec{y}$ the conditions appearing in \eqref{eq:kkt-sys} are equivalent to \eqref{eq:kkt-condition} by the change of variables $b = C$ and $a_i =  C - \deriv{f}{i}(\Vec{y})$.
\end{IEEEproof}
It is immediate to check that if \eqref{eq:kkt-condition} holds for some $\Vec{z}\in \Delta_n$ and some $C \in \bbR$, then
\begin{equation}
	C = \sum_{i = 1}^n z_{i}\deriv{f}{i}(\vec{z}).
\end{equation}
Here is a technical lemma that relates the limit points of trajectories with the multipliers, generalizing \cite[Prop.~3]{pelillo_torsello_payoff_monotonic_2006}.

\begin{lemma}\label{lem:mult}
    Let $\Omega$ be an open subset of $\bbR^n$, let $V$ be a closed non-empty subset of $\Omega$, let $f \in C^{1}(\Omega)$, and suppose that a function $\vec{z}(t)$ defined for $t \geq 0$ satisfies
    \begin{equation*}
        \begin{cases}        
            \dot{z}_{i} = z_{i} \left[ \deriv{f}{i}(\Vec{z}) 
        - \sum_{k = 1}^n z_{k } \deriv{f}{k}(\Vec{z}) \right],
        &i \in [n]\\
        \Vec{z}(0) \in V \cap \Delta_n\\
        \Vec{z}(t) \in V, &t \geq 0.\\
        \end{cases}
    \end{equation*}
    Then:
    \begin{itemize}
        \item $\Vec{z}(t) \in \Delta_n$ for all $t \geq 0$;
        \item there exists $\vec{z}^{(*)} = \lim_{t \to +\infty}\vec{z}(t) \in V \cap \Delta_n$;
        \item for every $i \in \supp{\vec{z}(0)} \setminus \supp{\vec{z}^{(*)}}$
    \begin{equation*}
        \deriv{f}{i}(\vec{z}^{(*)}) \leq \sum_{k = 1}^n z_{k}^{(*)}	\deriv{f}{k}(\vec{z}^{(*)}).
    \end{equation*}
    \end{itemize}
    \end{lemma}
\begin{IEEEproof}
    That $\Vec{z}(t) \in \Delta_n$ for all $t \geq 0$ is a consequence of the regularity condition together with the Picard-Lindel\"{o}f Theorem. It is easy to see that on $\Delta_n \cap \Omega$ the function $f(\Vec{z})$ is a Lyapunov function for the dynamics, arguing as in the proof of Theorem~\ref{thm:lyapunov}, hence, using also the compactness of $V\cap \Delta_n$, it follows that $z(t)$ admits a limit as $t \to +\infty$, which is necessarily an element of $V \cap \Delta_n$. What is left to prove is that $g_i(\vec{z}^{(*)}) \leq 0$ for every $i \in \supp{\vec{z}(0)} \setminus \supp{\vec{z}^{(*)}}$, where
    $g_i(\Vec{z})=  \deriv{f}{i}(\Vec{z}) - \sum_{k = 1}^n z_{k } \deriv{f}{k}(\Vec{z})$.
    By contradiction, suppose that $z_{i}^{(*)} = 0$ for some $i \in \supp{\vec{z}(0)}$ and that $g_i(\vec{z}^{(*)}) > 0$.
    The continuity of $g_i(\Vec{z})$ guarantees that some $T >0 $ exists such that $g_i(\vec{z}(t)) > 0$ for every $t \geq T$. Moreover, notice that $z_{i}(0) > 0$ by hypothesis, and $z_{i}(t) > 0$ by Theorem~\ref{thm:invariant} for all $t \geq 0$. This implies that $\dot{\vec{z}}_{i}(t) = z_{i}(t) g_{i}(\vec{z}(t)) > 0$ for all $t \geq T$, meaning that $z_{i}(t)$ is eventually increasing in $t$, and so $z_{i}^{(*)} = \lim_{t \to \infty} z_{i} (t) \geq z_{i}(T) > 0$, which is absurd.
\end{IEEEproof}

We also report here the information provided by Lyapunov's theorem \cite{helmke_optimization_1994} about our flows concerning the stability of the stationary points for \eqref{eq:ode-grad} and the relation with the maximization program \eqref{p:f}, similarly to what is mentioned in \cite{Bomze1997EvolutionTT}.
\begin{theorem}[Stability]\label{thm:stability}
Let $\vec{z} \in \Delta_{n}$ and consider the following properties:
\begin{itemize}
    \item[$(a)$] $\vec{z}$ is a stable stationary point for \eqref{eq:ode-grad};
    \item[$(b)$] $\vec{z}$ is an asymptotically stable stationary point for \eqref{eq:ode-grad};
    \item[$(c)$] $\vec{z}$ is an isolated stationary point for \eqref{eq:ode-grad};
    \item[$(d)$] $\vec{z}$ is a local solution of \eqref{p:f};
    \item[$(e)$] $\vec{z}$ is a strict local solution of \eqref{p:f};
\end{itemize}
then $[(a) \land (c) ] \iff  (b) \implies (e) \implies (a) \iff (d)$.
\end{theorem}
(
Theorem~\ref{thm:stability} has a simpler formulation whenever $f$ is a quadratic function, since then $(e)$ implies $(c)$, and consequently $(b)$ and $(e)$ are equivalent. Indeed, this equivalence appears in the replicator dynamics framework, in which the Lyapunov function is precisely a quadratic function \cite{hofbauer_evolutionary_1998}. That being said, we remark that this is not the case for the general $f$.%
    \footnote{It is not hard to imagine some $f$ having a strict global maximum $\vec{z}$ which is a limit point of a suitable sequence of stationary points, and in this case $\vec{z}$ is a strict local solution and hence a stable stationary point, even though lacking asymptotic stability.}

We consider as in Section~\ref{subsec:flow-for-i} the function
\begin{equation*}
    I(\Vec{z}) = \sum_{i = 1}^{n} c_i z_i - \sum_{j = 1}^{m} q_j \ln{q_j}
\end{equation*}
where
\begin{equation*}
    q_j = \prob{Y}(y_j) = \sum_{i = 1}^{n} p(j|i) z_i
\end{equation*}
and $c_i = \sum_{j = i}^{m} p(j|i) \ln[p(j|i)]$.
What follows are some properties regarding $I(\Vec{z})$.
\begin{proposition}
    The function $I(\Vec{z})$ is concave and continuous on $\Delta_n$. 
\end{proposition}
\begin{IEEEproof}
    Observe that for every $j$ the functions $q_j$ is a linear function in its variable $\Vec{z}$, and since the function $\phi(s)= -(s \ln{s})$ is concave and continuous on $[0,1]$, where we set $\phi(0)= 0$, then the composition $\phi \circ q_j = - q_j \ln{q_j}$ is also concave and continuous. Finally, $I(\Vec{z})$ is a sum of concave and continuous functions, and henceforth concave and continuous.   
\end{IEEEproof}

\begin{proposition}\label{prop:derivate}
    Let $\Vec{z} \in \bbR^n$ and suppose $\Vec{z} \geq \Vec{0}$.
    If $q_j(\Vec{z})\neq 0$ for every $j$, then
    \begin{itemize}
        \item $\deriv{I}{k}(\Vec{z}) = c_k - 1 -\sum_{j = 1}^{m} p(j|k)\ln{q_j}$,
    \end{itemize}
    Moreover, if $\vec{z} \in \Delta_n$, the following equality holds:
    \begin{itemize}
        \item $\sum_{k = 1}^n z_{k} \deriv{I}{k}(\Vec{z})
        =I(\Vec{z}) - 1 $.  
    \end{itemize}
\end{proposition}
\begin{IEEEproof}
    A direct computation shows that
    \begin{align*}
    \deriv{I}{k}(\Vec{z})
    &= c_k - \sum_{j = 1}^{m} \deriv{}{k}(q_j \ln{q_j})\\
    &= c_k - \sum_{j = 1}^{m} (\deriv{q_j}{k})(1 + \ln{q_j})\\
    &= c_k - \sum_{j = 1}^{m} p(j|k)(1 + \ln{q_j})\\
    &= c_k  -\sum_{j = 1}^{m} p(j|k) - \sum_{j = 1}^{m} p(j|k)\ln{q_j}\\
    &= c_k  -1 - \sum_{j = 1}^{m} p(j|k)\ln{q_j},
    \end{align*}
    and so
    \begin{align*}
        \sum_{k = 1}^n z_{k } \deriv{I}{k}(\Vec{z})
        &= \sum_{k = 1}^n z_{k } \left[c_k - 1 -\sum_{j = 1}^{m} p(j|k)\ln{q_j}\right]\\
        &= \sum_{k = 1}^n c_k z_{k}  - \sum_{j = 1}^{m} q_j \ln{q_j} - 1\\
        &=I(\Vec{z}) - 1
    \end{align*}
    using that $\sum_{k = 1}^n z_k = 1$ and $\sum_{k = 1}^n z_k p(j|k) = q_j$.
\end{IEEEproof}
The function $I(\Vec{z})$ can be not differentiable on some points on the boundary of $\Delta_n$, and in such points the Picard-Lindel\"{o}f's theorem cannot be applied. However, thanks to the following proposition, it is possible to extend by continuity the ODE associated with $I(\Vec{z})$ on the whole $\Delta_n$. For this extended ODE, the Peano's existence theorem \cite{birkhoff_rota_ODE} holds, thus allowing to solve \eqref{eq:ode-I-formal} in a genaralized sense also in the vicinity of points in which $I(\Vec{z})$ is not differentiable. Despite this, we are not able to prove the uniqueness of solutions in this case, and so it is unclear how to define the flow in the vicinity of such points.
\begin{proposition}
    For every $i \in [n]$, the function $z_i \deriv{I}{i}(\Vec{z})$, which is well defined on $\relint{\Delta_n}$, can be extended to a continuous function on $\Delta_n$.
\end{proposition}
\begin{IEEEproof}
    Let $\Vec{z} \in \relint{\Delta_n}$. Then
    \begin{align*}
            z_i \deriv{I}{i}(\Vec{z})
            &= z_i \left[c_i  -1 - \sum_{j = 1}^{m} p(j|i)\ln{q_j}\right]\\
            &=z_i (c_i -1)  - \sum_{j = 1}^{m} p(j|i) z_i\ln{q_j}\\
            &=z_i (c_i -1)  - \sum_{j : p(j|i)> 0} p(j|i) z_i\ln{q_j}.
    \end{align*}
    It is then sufficient to prove the continuity of the terms appearing in the summation.
    This is easily shown, in fact observe that for $p(j|i)> 0$ the function $p(j|i) z_i\ln{q_j}$ is continuous so long as $q_j >0$, and $p(j|i) z_i\ln{q_j} \to 0$ as $q_j \to 0^{+}$ due to the inequalities $0 \leq \lvert p(j|i) z_i\ln{q_j}) \rvert \leq \lvert q_j \ln q_j \rvert$.
\end{IEEEproof}

%% file: sections/implementation.tex
\noindent By what we have seen in Section~\ref{sec:contribution-i-specific}, the continuous-time dynamics initialized $\relint{\Delta_n}$ is always confined in $\Delta_n$. However, this may not be the case for discrete dynamics obtained via numerical methods. In fact, this fails to happen also in the case of a mere application of the Euler method. In the preliminary experiments we have run, we have found that the discrete trajectory can happen to jump to a vector $\vec{z} \in \bbR^{n}$ having $\sum_{i=1}^n z_i \neq 1$ or $ z_i \notin [0, 1]$ for some $i$, causing the dynamics to blow-up in the subsequent iterations. 
This indeed is not a surprise in case the step size is not sufficiently small. In fact, it is quite reasonable that moving for an excessive distance in the direction prescribed by the ODE could result in exiting $\Delta_n$.
However, we also observed this phenomenon for small values of the step size, together with a violation of the constraint $\sum_{i=1}^n z_i = 1$. Theoretically, this should not happen, unless the dynamics exits the positive orthant first. This is a consequence of the regularity condition.
\begin{proposition}
    Let $\Vec{z}^{(0)}\in \relint{\Delta_n}$, and let $\Vec{z}^{(1)}$ be obtained by applying to $\Vec{z}^{(0)}$ one step of Euler method that approximates the flow of \eqref{eq:ode-I-formal}. Then  $\sum_{i = 1}^n z_i^{(1)} = 1$, regardless of the step size used.
\end{proposition}
\begin{IEEEproof}
    Set $g_i = z_i^{(0)} [\deriv{I}{i}(\Vec{z}^{(0)}) - \sum_{k = 1}^{n} z_i^{(0)} \deriv{I}{i}(\Vec{z}^{(0)})]$ for every $i \in [n]$ and call
    $\tau$ the step size used to define $\Vec{z}^{(1)}$. By construction
    \begin{align*}
            z_i^{(1)} &= z_{i}^{(0)} + \tau z_{i}^{(0)}g_i, & i \in [n].
    \end{align*}
    Since $\Vec{z}^{(0)}\in \Delta_n$, then $\sum_{i = 1}^n z_{i}^{(0)} = 1$, and so \begin{equation*}
    \sum_{i = 1}^n z_i^{(1)}
    = \sum_{i = 1}^n z_{i}^{(0)} + \tau  \sum_{i = 1}^n z_{i}^{(0)} g_i = 1
\end{equation*}
    using that $\sum_{i = 1}^n z_i^{(0)} g_i = 0$ by the regularity condition.
\end{IEEEproof}
We believe that this phenomenon is a consequence of the floating point arithmetic: our opinion is that the small truncation errors occurring at every step of the numerical method have a cumulative effect that is not negligible, and that leads the dynamics to depart from the simplex.
Indeed, increasing the machine precision reduces the magnitude of this phenomenon, and this supports the interpretation we gave.

\begin{figure}[ht!]
     \centering
     \subfloat[]{\includegraphics[width=0.49\linewidth]{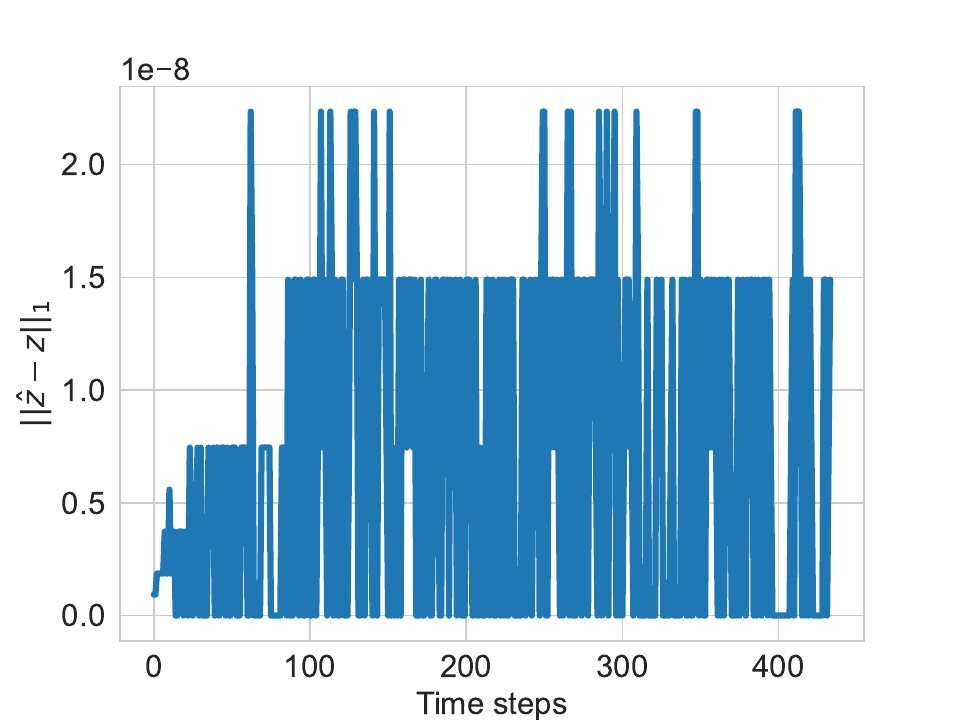}}
     \hfill
     \subfloat[]{\includegraphics[width=0.49\linewidth]{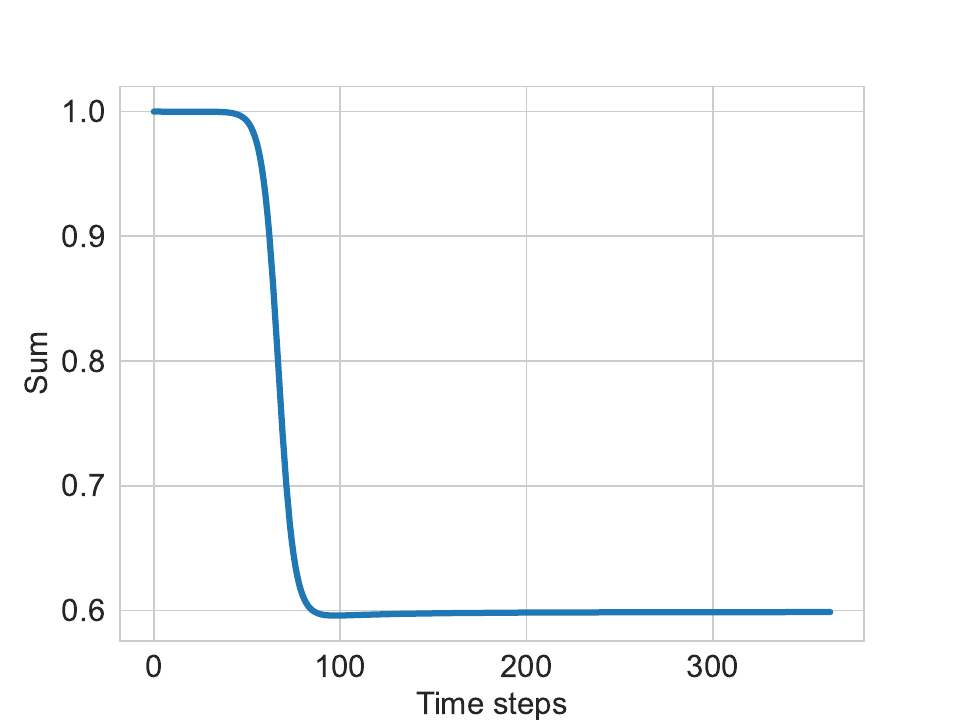}}
     \caption{Effect of the normalization procedure for $\tau = 1$. Using the adjusted Euler method (a), the $L^1$ approximation error of $\vec{z}^{(k)}$ in replacing it with its normalized projection $\hat{\vec{z}}^{(k)}$ is negligible. Using the classic Euler method (b), the sum of the components in $\vec{z} ^{(k)}$ depart from the value $1$, in contrast with the theoretical behavior.}
    \label{fig:diffAtTimeStep}
\end{figure}

Overcoming this issue has motivated the adoption of the adjusted Euler method in the experiments, as mentioned in Section~\ref{sec:experiments}.
The additional normalization procedure --- essentially a projection over $\Delta_{n}$ --- clearly solves the issue, and for small values of the step size, the perturbation introduced in this way is very small at each iteration
Figure \ref{fig:diffAtTimeStep} shows an example of a channel where the vector flow is discretized with the adjusted Euler method and with the classical one. In Figure \ref{fig:diffAtTimeStep} (a), obtained with the adjusted Euler method, we can observe how small is the the $L^1$ approximation error of $\vec{z}^{(k)}$ in replacing it with its normalized projection $\hat{\vec{z}}^{(k)}$ at every iteration, being in the order of $10^{-8}$. In contrast, Figure~\ref{fig:diffAtTimeStep} (b), obtained with the classic Euler method, shows how the value $\sum_{i = 1}^n z_i^{(k)}$ moves away increasingly from the value $1$ as $k$ increases. 

